\documentclass{elsart}
\usepackage{graphicx,amssymb}
\usepackage{multicol}
\usepackage[titles]{tocloft}
\journal{Physics Reports}
\begin{document}

\begin{frontmatter}

\title{The Thermodynamic Model for Nuclear Multifragmentation}

\author[a,b]{C. B. Das}
\author[a]{S. Das Gupta{\thanksref{cor1}}}
\author[c]{W. G. Lynch}
\author[d]{A. Z. Mekjian}
\author[c]{M. B. Tsang{\thanksref{cor2}}}
\address[a]{Physics Department, McGill University,
Montr{\'e}al, Canada H3A 2T8}
\address[b]{Physics Division, Variable Energy Cyclotron Centre,
Kolkata 700 064, India}
\address[c]{National Superconducting Cyclotron Laboratory 
and Physics and Astronomy Department,
Michigan State University, East Lansing, MI 48824, USA}
\address[d]{Physics Deptartment, Rutgers University, Piscataway,
NJ 08854, USA}

\thanks[cor1]{E-mail: dasgupta@physics.mcgill.ca}
\thanks[cor2]{E-mail: tsang@nscl.msu.edu}

\begin{minipage}{\textwidth}
\tableofcontents

\listoffigures

\begin{abstract}
A great many observables seen in intermediate energy heavy ion collisions
can be explained on the basis of statistical equilibrium.
Calculations based on statistical equilibrium can be implemented in
microcanonical ensemble (energy and number of particles
in the system are kept fixed), canonical ensemble (temperature and
number of particles are kept fixed) or grand canonical ensemble
(fixed temperature and a variable number of particles but with an assigned
average).  This paper deals with calculations with canonical ensembles.
A recursive relation developed recently allows calculations with
arbitrary precision for many nuclear problems.  Calculations are
done to study the nature of phase transition in intermediate energy
heavy ion collision, to study the caloric curves for nuclei
and to explore the possibility of negative specific heat
because of the finiteness of nuclear systems.   The model can also be used
for detailed calculations of other observables not connected with
phase transitions, such as populations of selected isotopes in a
heavy ion collision.

The model also serves a pedagogical purpose.  For the problems
at hand, both the canonical and grand canonical solutions are
obtainable with arbitrary accuracy hence we can compare the
values of observables obtained from the canonical calculations with
those from the grand canonical.
Sometimes, very interesting discrepancies are found.

To illustrate the predictive power of the model, calculated observables are compared with  
data from the central collisions of Sn isotopes.

\end{abstract}

\end{minipage}

\end{frontmatter}

\newpage

\section{Introduction}
Consider a central collision of two heavy ions.  Nucleons from one
nucleus will collide with nucleons from another nucleus. After a
few collisions a given nucleon may lose the identity of its
source.  The system then more resembles a hot fluid of nucleons
in an overall volume.  Depending upon the original beam energy,
this system may undergo an initial compression and then begins to
decompress.  During this time the nucleons will interact with each
other, at least between the nearest neighbours. As the density of
the system decreases, higher density regions will develop into
composites. As this collection of nucleons begin to move outwards,
rearrangements, mass transfers, nucleon coalescence and most
physics will continue to happen until the density decreases so much that
the mean free paths for such processes become larger than the
dimension of the system. Subsequently the objects follow the 
long range coulomb
trajectories. Our objective is to have a soluble model which
describes the physics of the situation at this freezeout density
when one averages many nucleus-nucleus collisions.

Although we chose central collisions to describe this scenario a similar
situation will arise even for semi-central or semi-peripheral collisions.
In such cases one may have projectile like fragment (and target like
fragment and participants, region of violent collisions).  For example, a
projectile fragment may be excited which resembles a system of hot particles
whose centre of mass velocity is close to that of the projectile
\cite{Hubele}.

The central assumption of the present article (and many others) is
that equilibrium statistical mechanics can be used to describe the
hot fluid of nucleons.  Even the most well prepared experimental
measurement of an energetic $nucleus-nucleus$ collision represents
an average of a very large number of initial states. In addition
to this large number of different initial states,  a large
number of $nucleon-nucleon$ collisions occur within each
$nucleus-nucleus$ collision. Together, this means that for many
experimental observables almost all the relevant phase-space can be
opened up and described by the microcanonical ensemble in which
the probability of reaching a channel $y$ is
$\Omega(y)/\sum_y\Omega(y)$. Here $\Omega(y)$ is the phase-space
volume in the channel $y$. In the canonical ensemble the
corresponding expression \cite{Reif} is writen as:
$\exp(-f(y)/T)/\sum\exp(-f(y)/T)$.  Here $f(y)$ is the free energy
in the channel $y$.  Since $f(y)=-TlnQ(y)$ where $Q(y)$ is the
canonical partition function in the channel $y$, an equivalent
expression is $Q(y)/\sum Q(y)$. A more detailed discussion of
statistical equilibrium using reaction rates is given in Appendix
A.

The obvious experimental observables in heavy ion collisions are the
number of nucleons and composites and their velocity distributions
that result after the collision.  The calculation of these in
equilibrium statistical mechanics for Bevalac physics is more than
twenty five years old \cite{Mekjian1,Gosset,Dasgupta1}. At that time,
the grand canonical ensemble was used to describe data from the  Bevalac 
which normally
used beam energies higher than 250 MeV/nucleon.  However, at 
these energies most of the subtle
and interesting features of equilibrium statistical mechanics as
it pertains to heavy ion collision have disappeared.  
As the cross-sections of composites fall rapidly with $A$,
the mass number,  the most interesting results were productions of
new particles such as pions and kaons, which can be included in the
statistical model.  Some discussion of this production is also
given in Appendices A and C. 
Even so, Bevalac experiments brought out beautiful
features of dynamics and established narrow limits on
compressibility of nuclear matter and the momentum dependence
of the real part of the optical potential.

The applications of equilibrium statistical mechanics for intermediate
energy heavy ion collisions started in the eighties.
At these energies,  the efforts switched to microcanonical
ensembles although the concept of temperature was sometimes used
\cite{Randrup,Gross1,Bondorf}. One model called the Copenhagen SMM
(statistical multifragmentation model) is frequently used \cite{Bondorf}.
Another popular model is the Berlin model \cite{Gross1}. The use of
the canonical ensemble,
the main topic of this paper, is more recent\cite{Dasgupta2}.
It is as easy to implement as the grand canonical (and more
accurate since fluctuations in the number of particles are
eliminated: these sometimes cause large errors in computations of
observables).  It is orders of magnitudes simpler than the microcanonical
ensemble although in the latter more fine tuning can be done.  These
fine tunings do not appear important for most observables.

What are the important issues we want to learn about in intermediate
energy heavy ion collisions?  For many, it is to extract from data
signals of a liquid-gas
phase transition in nuclear matter.  Nuclear matter
is a hypothetical large chunk of matter with $N=Z$ where the coulomb
interaction has been switched off.  The $p-V$ diagram for nuclear
matter with reasonable forces looks like a Van der Waals equation
of state \cite{Jaqamann}.  One would then expect to see a liquid-gas
phase transition if the experimental conditions are optimal.  Such
optimal conditions are discussed by Curtin, Toki and Scott
\cite{Curtin} and Bertsch and Siemens \cite{Bertsch1}.  For Bevalac
energies the evolution of the temperature would go above the phase
transition temperature but accelerators at the National Superconducting
Cyclotron Laboratory (NSCL), the Texas A$\&$M cyclotron,
the Grand Accelerateur National D'ions
Lourds (GANIL) and at Gesellschaft fur
Schwerionenforschung mbH (GSI)
can reach the liquid-gas phase transition region and offer the best
possibility for experimental study.  Further details of theoretical
considerations which prompt an experimental investigation of the
liquid-gas phase transition can be found in \cite{Dasgupta3}.

Unfortunately, this investigation of liquid-gas phase transition
in intermediate energy heavy ion collisions is fraught with many
difficulties. Phase transitions occur in very large systems.  In
nuclear collisions, we are limited to three to four hundred
nucleons (sometimes much less). For finite systems, signals of
phase transition get diluted and distinctions between
first order and second order transitions get blurred.  The coulomb
interaction, which prevents large nuclei from forming also
interferes with the signals.  It is thus necessary to use
theories to clarify the situation.  If
one has a theory which fits many data, not necessarily 
related to phase transitions, but in addition predicts a
phase transition one has some hope for the model to be valid. In
this paper we will discuss phase transitions but in addition, 
data which will be compared to the thermodynamic model
predictions.

\section{The Basic Formulae}
This section sets up the basic formulae of the model
\cite{Dasgupta2,Bhat1}.

If there are $A$ identical particles of only one kind in an enclosure
at temperature $T$, the partition function of the system
can be written as
\begin{eqnarray}
Q_A=\frac{1}{A!}(\omega)^A
\end{eqnarray}
Here $\omega$ is the partition function of one particle.  For a spinless
particle without any internal structure
$\omega=\frac{V}{h^3}(2\pi mT)^{3/2}$; $m$ is the mass
of the particle; $V$ is the available volume within which each particle
moves; $A!$ corrects for Gibb's paradox.  If there are many
species, the generalisation is
\begin{eqnarray}
Q_A=\sum\prod_i\frac{(\omega_i)^{n_i}}{n_i!}
\end{eqnarray}
Here $\omega_i$ is the partition function of a composite which has $i$
nucleons.  For a dimer $i=2$, for a trimer $i=3$ etc.  Eq. (2) is
no longer trivial to calculate.  The trouble is with the sum in the right hand
side of eq. (2).  The sum is restrictive.  We need to consider only
those partitions of the number $A$ which satisfy $A=\sum in_i$.  The number
of partitions which satisfies the sum is enormous.  We can call a
given allowed partition to be a channel.  The probablity of the occurrence
of a given channel $P(\vec n)\equiv P(n_1,n_2,n_3....)$ is
\begin{eqnarray}
P(\vec n)=\frac{1}{Q_A}\prod\frac{(\omega_i)^{n_i}}{n_i!}
\end{eqnarray}
The average number of composites of $i$ nucleons is easily seen from
the above equation to be
\begin{eqnarray}
\langle n_i \rangle = \omega_i\frac{Q_{A-i}}{Q_A}
\end{eqnarray}
Since $\sum in_i=A$, one readily arrives at a recursion relation
\cite{Chase}
\begin{eqnarray}
Q_A=\frac{1}{A}\sum_{k=1}^{A}k\omega_kQ_{A-k}
\end{eqnarray}
For one kind of particle, $Q_A$ above is easily evaluated on a computer for
$A$ as large as 3000 in matter of seconds.  It is this recursion relation
that makes the computation so easy in the model.  Of course, once one has
the partition function all relevant thermodynamic quantities can be
computed.

We now need an expression for $\omega_k$ which can mimic the nuclear physics
situation.  We take
\begin{eqnarray}
\omega_k=\frac{V}{h^3}(2\pi mT)^{3/2}\times q_k
\end{eqnarray}
where the first part arises from the centre of mass motion of the
composite which has $k$ nucleons and $q_k$ is the internal
partition function. For $k=1$, $q_k=1$ and for $k\ge 2$ it is
taken from the Fermi-gas model. For each composite consisting of
$k$ nucleons, we approximate the intrinsic free energy at freeze-out
by $E-TS=-W_0k+\sigma(T)k^{2/3}+kT^2/\epsilon_0-T\times
2kT/\epsilon_0$ where $\epsilon_0$ is a constant.  This gives
\begin{eqnarray}
q_k=\exp[(W_0k-\sigma (T)k^{2/3}+T^2k/\epsilon _0)/T]
\end{eqnarray}
Here, as in \cite{Bondorf}, $W_0$=16 MeV is the volume energy term,
$\sigma(T)$
is a temperature dependent surface tension term.  The value of
$\epsilon_0$ is taken to be 16 MeV.  The explicit expression for $\sigma(T)$
used here, as in \cite{Bondorf}, is 

$\sigma(T)=\sigma_0[(T_c^2-T^2)/(T_c^2+T^2)]^{5/4}$ 

with $\sigma_0=$18 MeV and $T_c=18$ MeV.  In the nuclear case one might be
tempted to interpret $V$ of eq.(6) as simply the freeze-out volume but
it is clearly less than that; $V$ is the volume available to the particles
for the centre of mass motion.  Assume that the only interaction between
clusters is that they can not overlap one another.  Then
in the Van der Waals spirit we take
$V=V_{freeze}-V_{ex}$ where $V_{ex}$ is taken here to be constant and
equal to $V_0=A/\rho_0$.  The assumption that the interaction between
different composites is only reflected through an excluded volume
and that this excluded volume is independent of multiplicity is an
idealisation which will fail for a non-dilute system.  We therefore
restrict the model, somewhat arbitrarily to volumes $V_{freeze}\ge 2V_0$.
There are experimental signatures that $V_{freeze}$ is indeed greater
than $2V_0$ \cite{Dasgupta3} so this is not a debilitating feature
of the model.  In all our considerations we restrict $\rho/\rho_0$
to less than 0.5.

Among quantities of interest is the inclusive cross-section 
given by eq.(4).  Actually this is a simplification.  The
occupation given by eq.(4) is the occupation of the composite with
$i$ nucleons at temperature $T$.  Both the ground state and the
excited states contribute to $\langle n_i \rangle$.  Some of the
excited states will be particle unstable and will decay into lower
mass composites before they reach the detector.  On the other
hand, some higher mass composites, will, by the same argument,
decay into the composite $i$.  In later sections, where we compare
populations with data, this aspect will be taken care of.
The expression for $E$ at a given temperature $T$ is simple (this
is needed for a caloric curve which is measured in experiments).
The energy carried by one composite is given by 

$E_k=T^2\partial
ln\omega_k/\partial T=\frac{3}{2}T+k(-W_0+T^2/\epsilon_0)
+\sigma(T)k^{2/3}-T[\partial\sigma(T)/\partial T]k^{2/3}$.  

Of these the first term comes from the centre of mass motion and the rest from
$q_k$. The term $T[\partial\sigma(T)/\partial T]k^{2/3}$ comes
from the temperature dependence of the surface tension.  It has a
small effect. The energy of the whole system is given by
$E=T^2\frac{1}{Q_A} \frac{\partial Q_A}{\partial T}$.  Using
eqs.(2) and (4) we arrive at a very transparent formula :$ E=\sum
\langle n_k \rangle E_k$.  The pressure is given by $p=T\partial
lnQ_A/\partial V$.  If for  purposes of illustration, we neglect
the long range Coulomb interactions and use eqs.(2) and (4) we get
$p=T\frac{1}{V}\sum \langle n_i \rangle$. This is just the law of
partial pressures.

For analysing phase transitions in the model, it is
very useful to calculate the average value of the largest
cluster in the ensemble. Eq. (2) shows that the size of the largest
cluster varies.  In that ensemble there is a term $\frac{\omega_1^A}
{A!}.$  For this the largest cluster is the monomer.
For example in eq.(2) we also have a term
$\frac{\omega_1^n}{n!}\frac{\omega_2^{(A/2-n/2)}}{(A/2-n/2)!}.$  Here the
largest cluster is the dimer.  Consider building $Q_A$ with
$\omega_1,\omega_2......\omega_k,0,0,0,0..$.  In this ensemble
the largest cluster will span from monomer upto a composite with
$k$ nucleons.  Let us label this partition function $Q_A(\omega_1,
\omega_2,....\omega_k,0,0,0..)$.  Let us also build a $Q_A$
where the largest non-zero $\omega$ is $\omega_{k-1}$.  The partition function
is $Q_A(\omega_1,\omega_2,..\omega_{k-1},0,0,0,0)$.  In this
ensemble all the previous channels are included except where the
largest cluster had $k$ nucleons.  If we define
\begin{eqnarray}
\Delta Q_A(k)=Q_A(\omega_1,\omega_2...\omega_k,0,0,..)-Q_A(\omega_1,\omega_2
...\omega_{k-1},0,0....), \nonumber
\end{eqnarray}
then the probability of the largest cluster having
$k$ nucleons is
\begin{eqnarray}
Pr(k)=\frac{\Delta Q_A(k)}{Q_A(\omega_1,\omega_2......\omega_A)}
\end{eqnarray}
If we now label the average value of the largest cluster as
$\langle k_{max} \rangle$,
then $\langle k_{max} \rangle =\sum k\times Pr(k)$.  A more useful quantity is
$\frac{\langle k_{max} \rangle}{A}$.  The limits of this are $\approx 0$ and 1.

Another interesting quantity which has been the subject of an enormous
amount of interpretation \cite{Moretto} is the multiplicity distribution
of a species or a group of species.  In most models this requires
a very elaborate Monte-Carlo calculation.  In the canonical ensemble
there is an elegant equation.
\begin{eqnarray}
P_n(k)=\frac{1}{Q_A}\frac{\omega_k^n}{n!}Q_{A-nk}(\omega_1,\omega_2...
\omega_{k-1},\omega_k=0,\omega_{k+1}..\omega_A)
\end{eqnarray}
Here $P_n(k)$ is the probability of obtaining the composite $k$ n times.

The strength of the canonical model as described here lies in the fact
that all calculations above avoid Monte-Carlo sampling.  In many other models,
a Monte-Carlo sampling over the channels is required.  Since the number
of channels is enormous, this requires great ingenuity but also much
more computer time.

The model of one kind of particles in which composites have a volume energy,
a surface energy and excited states is already very useful for
investigations of phase transition, caloric curves etc. and we will
pursue this in latter sections a great deal.  Let us nonetheless
introduce here the model
with two kinds of particles (so that one can compare with actual
nuclear cases)  \cite{Dasgupta3,Bhat1,Souza1}.
Now a composite is labelled
by two indices $\omega\rightarrow \omega_{i,j}$.  The partition function
for a system with $Z$ protons and $N$ neutrons is given by
\begin{eqnarray}
Q_{Z,N}=\sum\prod_{i,j}\frac{\omega_{i,j}^{n_{i,j}}}{n_{i,j}!}
\end{eqnarray}

There are two constraints: $Z=\sum i\times n_{i,j}$ and
$N=\sum j\times n_{i,j}$.
These lead to two recursion relations any one of which can be used.
For example
\begin{eqnarray}
Q_{Z,N}=\frac{1}{Z}\sum_{i,j}i\omega_{i,j}Q_{Z-i,N-j}
\end{eqnarray}
where
\begin{eqnarray}
\omega_{i,j}=\frac{V}{h^3}(2\pi mT)^{3/2}(i+j)^{3/2}\times q_{i,j}
\end{eqnarray}
Here $q_{i,j}$ is the internal partition function.  These could be taken
from experimental binding energies, excited states and some model for the
continuum or from the liquid drop model in combination with other models.
The versatility of the method lies in being able to accommodate any choice
for $q_{i,j}$.  A choice of $q_{i,j}$ from a combination of the liquid drop
model for binding energies and the Fermi-gas model for excited states
that has been used is
\begin{eqnarray}
q_{i,j}=\exp\frac{1}{T}[W_0a-\sigma a^{2/3}-\kappa\frac{i^2}
{a^{1/3}}-s\frac{(i-j)^2}{a}+T^2a/\epsilon_0]
\end{eqnarray}
where $a=i+j, W_0=15.8$ MeV, $\sigma=18.0$ MeV, $\kappa=0.72$ MeV,
$s=23.5$ MeV
and $\epsilon_0=16$ MeV.
One can recognise in the parametrisation above, the volume term, the
surface tension term, the coulomb energy term, the symmetry energy term
and contributions from excited states.

The coulomb interaction is long range.  Some effects of the coulomb interaction
between different composites can be included in an approximation
called the Wigner-Seitz approximation.  We assume, as usual, that
the break up into different composites occurs at a radius $R_c$, which
is greater than the normal radius $R_0$.  Considering this as a process
in which a uniform dilute charge distribution within radius $R_c$
collapses successively into denser blobs of proper radius $R_{i,j}$,
we write the coulomb energy as \cite{Bondorf}
\begin{eqnarray}
E_C=\frac{3}{5}\frac{Z^2e^2}{R_c}+\sum_{i,j}\frac{3}{5}\frac{i^2e^2}
{R_{i,j}}(1-R_0/R_c)
\end{eqnarray}
It is seen that the expression is correct in two extreme limits:
very large freeze-out volume ($R_c\rightarrow\infty)$ or if the
freeze out volume is the normal nuclear volume so that one has
just one nucleus with the proper radius.

For the thermodynamic model that we have been pursuing, the constant
term $\frac{3}{5}\frac{Z^2e^2}{R_c}$ is of no significance since
the freeze-out volume is assumed to be constant.  In a mean-field
sense then one would just replace the coulomb term in eq.(13)
by $\kappa\frac{i^2}{a^{1/3}}(1.0-(\rho/\rho_0)^{1/3})$.

Before we leave this section, we mention that the mass parametrisation
implied by eq.(13) can be vastly improved with only slight complications.
We will later present results with the improved formula \cite{Souza1}.
A pedagogical issue:
although we have derived results here based on eq.(1) which takes
care of (anti)symmetrisation only approximately it is shown
in \cite{rab5} that the specific structure of
eqs.(5) and (11) occur more generally when (anti)symmetrisation is
included properly.  Part of this argument is presented in appendix B
which also demonstrates that results based on this section
are quite accurate.

\section{General features of yields of composites}
We pursue here the model of one kind of particles.  For 200 particles
at a constant freeze-out volume =$3.7V_0$ we have plotted
in fig.1 $\langle n_k \rangle$ (in
the figure we call this $Y(a)$=yield of composite of mass $a$) at
three temperatures.  At the lowest temperature shown the curve has
a $U$ shape.  The yields $Y(a)$ first begin to fall, then reach
a minimum and then the yields for heavier masses increase finally
cutting off at 200.  In the literature the heavy fragments are called the
liquid phase.  The light fragments are gas particles.  As the temperature
increases the maximum at the higher $a$ decreases in height finally
disappearing at $\sim$6.35 MeV.  At higher temperature $Y(a)$ falls
monotonically.
The surface tension plays a crucial role in this evolution.  At any
temperature the lowest value of the free energy $E-TS$ will be obtained.
It costs in the energy term $E$ to break up a system.  A nucleus of $A$
nucleons has less surface than the total surface of two nuclei each of $A/2$
nucleons (the volume energy term has no preference between the two
alternatives).  Therefore at low temperature one will see a large chunk.
The -$TS$ term favours break up into small objects.  The competition
between these two effects leads to the general features of $Y(a)$
as a function of temperature.  As we will see in the next section,
the temperature at which the maximum of the yield at the high side of
$a$ disappears is the phase transition temperature.

Similar features are seen also in other models of multifragmentation
as applied to nuclear physics.  The earliest such model was the
percolation model \cite{Bauer1,Campi1}.  The model has a parameter
$p$ which gives the probability of two nearest neighbour sites
joining together as in a composite.  Beyond a certain value of $p$,
a percolating cluster is formed which goes from edge to edge of the
system.  This corresponds to the large cluster which forms at the
lower temperature in fig.1.  The lattice gas model \cite {Pan1} has
similarity with the percolation model but has a Hamiltonian, includes
percolation model as a subset \cite{Dasgupta4} and also includes
the formation of a percolating cluster

\section{Phase transition in the model }

\subsection{Signatures from thermodynamic variables}

We now begin the discussion of a phase transition in the model.  The
free energy of a system of particles is given by $F=-T$ln$Q_A$ and
ln$Q_A$ is directly calculable with eq.(5).  For a system of 200
and 2000 particles, the free energy per particle is shown in
the top panel of fig.2, as a function of 
temperature for fixed freeze-out density 0.27$\rho_0$.  An approximate break in
the first derivative of $F/A$ is seen to develop at $\approx$6.35
MeV for 200 particles and at $\approx$7.15 MeV for 2000 particles.
We believe the break would be rigorous if we could go to an infinite system.
A break in the first derivative implies a first order phase transition
and a discontinuous change in the value of entropy per particle.
This would imply that the specific heat at constant volume per particle
$c_v=(\frac{\partial(E/A)}{\partial T})_V$ would go
through a peak (for an infinite system this peak would go to
$\infty$).  We show this in the midddle panel of fig.2 for systems
of 200 and 2000
particles, where we find the width of the peak decreases and the height of
the peak increases as the particle number increases.  As 
expected, the temperature where the specific heat maximises also
coincides with the temperature at which the maximum in the high side
of $a$ (fig.1) just disappears.

\vskip 0.2in
\begin{center}
\includegraphics[width=3.5in,height=3.5in]{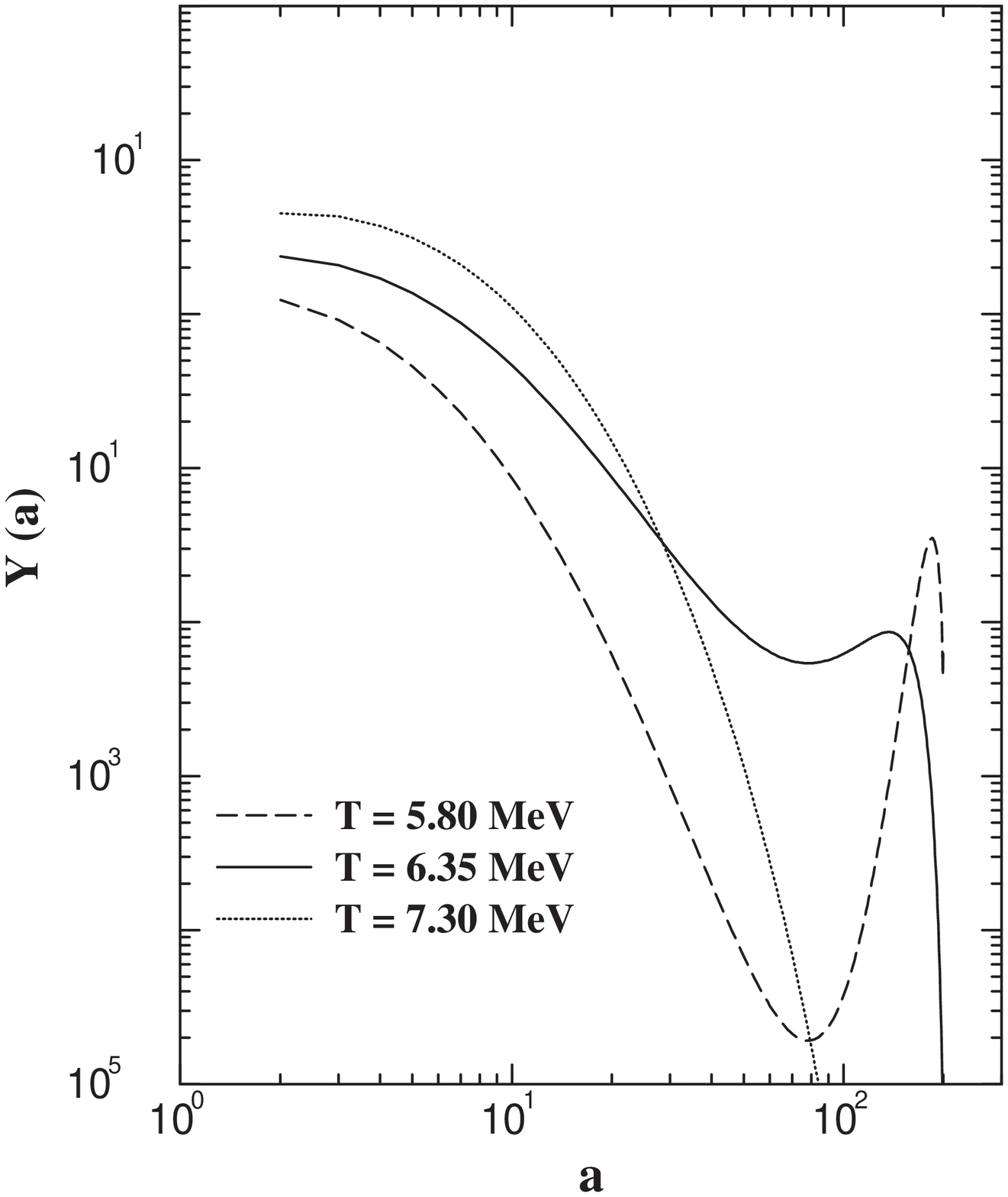}
\end{center}

Fig.1 Yield $Y(a)$ against $a$ at three different temperatures.  The 
dissociating system has 200 particles and the freeze-out density is 
$0.27\rho_0$.

\vskip 0.2in
Another very interesting quantity is the quantity $\langle k_{max} \rangle /A$
(i.e. the size of the largest cluster) as
the temperature varies.  This can be calculated using eq.(8).We
define $T_b$ as the temperature where the break in the derivative
of the free energy occurs (this is the first order phase transition
temperature). Calculating the size of the largest cluster at different
temperatures, we find that $\langle k_{max} \rangle /A$ approaches 1 as $T < T_b$
and approaches a small number as $T > T_b$.  The change is smooth
for low mass nuclei (bottom panel of fig.2) but becomes more sudden for larger
systems.  For large systems there is a large blob (i.e., liquid)
below $T_b$ which disappears as soon as $T$ crosses $T_b$.  This
we think is a very engaging example of boiling emerging from a
theoretical calculation.

\vskip 0.2in
\begin{center}
\includegraphics[width=3.5in,height=3.5in]{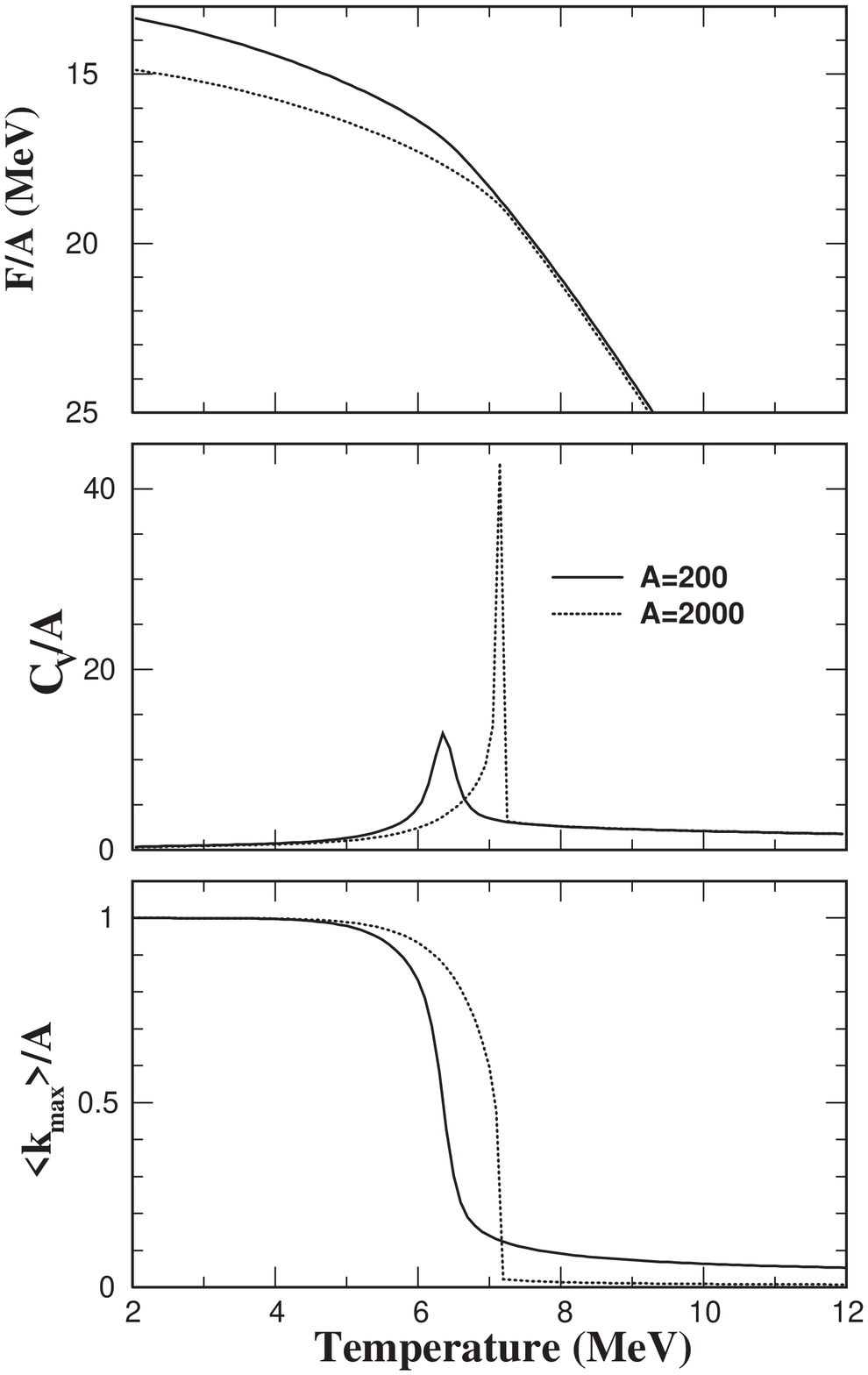}
\end{center}

Fig. 2. The free-energy per particle, the specific
heat at constant volume, $C_V/A$ and the size of the largest cluster
as a function of temperature, systems of 200 and 2000 particles.

\vskip 0.2in
To summarise, the thermodynamic model predicts unequivocally
a first order phase transition at intermediate energy.  In the
realm of density $\rho/\rho_0\le 0.5$ for which we believe
the model to be reasonable there is no critical point.  Bugaev
et al. \cite{Bugaev} have taken the model beyond this range of
density and find that the critical density is $\rho/\rho_0=1$
and the temperature is 18 MeV when the surface tension
$\sigma(T)$ goes to zero.  We end this section by noting that
microcanonical calculations using statistical equilibrium were
also suggestive of a first order phase transition occurring at
intermediate energy \cite{Bondorf2,Randrup}.

\subsection{Power-law and scaling behaviour of composite yields}

A rather large part of literature in heavy ion reaction
postulates that in multifragmentation at
intermediate energy one is near the critical point of nuclear matter.
One then proceeds to determine from the data the critical temperature
and various critical exponents.  The working formula, obtained from
models of critical phenomena (to see how the formula arises, refer to
\cite{Stauffer,Widom}) is
\begin{eqnarray}
\langle n_a \rangle = a^{-\tau}f(a^{\sigma}(T-T_c))
\end{eqnarray}
Here $\tau$ is called the Fisher exponent \cite{Fisher}, $a$ is the mass
number of the composite, $\sigma$ is a critical exponent and $T_c$
the critical temperature ; $f$ is as yet an unspecified function but
instead of being a general function of $a$ and $T$ it is a function only
of the combination $a^{\sigma}(T-T_c)$.  This is called scaling.
At $T=T_c$ the yield $\langle n_a \rangle = a^{-\tau}f(0)$ is a pure power law but
away from $T_c$ it will deviate from a power law.

In intermediate energy collisions, even if we proceed under the
assumption that one is observing critical phenomena we can not
expect near perfect fit to eq.(15) whose validity depends upon
the dissociating system being very large.  Also the range of $a$
is to be chosen judiciously.  It can not be very small (since eq.(15)
applies to ``large'' $a$'s \cite{Stauffer}).  But $a$ also should
be truncated on the high side (significantly smaller than the size
of the dissociating system).

With these provisos we can at best expect a moderately good fit.
Extracting $\tau,\sigma$ and $T_c$ from a given set of $\langle n_a \rangle$
(either from experiment or models)
when only an approximate fit is
expected is non-trivial and not unique.  We skip the details here
which are given in \cite {Elliott1,Elliott2,Scharenberg}.  A
more sophisticated method of extraction of the relevant
parameters can be found in \cite{Gulminelli1}.  The same technique
is used in \cite{Das1}.

The EOS collaboration \cite{Gilkes} obtained data from break up
of 1.0 GeV per nucleon gold nuclei on a carbon target.  Depending
upon the impact parameter, the excitation energy (or the temperature)
of the projectile like fragment which breaks up into
many composites will vary.  In \cite{Elliott1,Elliott2} it is argued that
$T$ in eq.(15) varies linearly with the charged multiplicity $m$
and the scaling function of eq.(15) is changed from
$f(a^{\sigma}(T-T_c))$ to $f(a^{\sigma}(\frac{m-m_c}{m_c}))$.
Here $m_c$ is the critical multiplicity.  Having determined from the
data $\tau,\sigma$ and $m_c$ (as mentioned before we are skipping
the details of how the extraction is done but this can be found
in \cite{Elliott1,Elliott2}) one then verifies if the scaling law
works: that is, we check if for all $a$'s,  $\langle n_a \rangle a^{\tau}$ will
fall on the same curve
when plotted as a function of $a^{\sigma}(m-m_c)/m_c$.  How well this
works can be seen, for example, in fig.(18) of \cite{Scharenberg}.
The deviations from the hypothetical ``universal'' curve are by
no means negligible but can we assume that the scatter of points
is entirely finite particle number effect and conclude
that we have indeed seen evidence of critical phenomena?

To resolve this, we play a theoretical game.  We take the thermodynamic model
(which we know has only a first order phase transition), pick a system
with particle number $A$, generate $\langle n_a \rangle$
for different temperatures $T$ and from these data extract best possible
values of $\tau,\sigma$ and $T_c$. Having obtained these we examine
how well the scaling law
applies.  This is shown in fig.3.  The figure is taken from \cite{Das1}
where other similar examples are displayed.  For the model, deviations
from one ``universal'' curve are smaller than what the EOS 
collaboration data gave.  We
might conclude we have extracted the model critical temperature and the
critical exponents.  These would be wrong conclusions of course because
the model has only a first order phase transition.  In fact the value of
$T_c$ one extracts this way is quite close to $T_b$, the first order
phase transition temperature.

\vskip 0.2in
\begin{center}
\includegraphics[width=4.5in,height=4.5in]{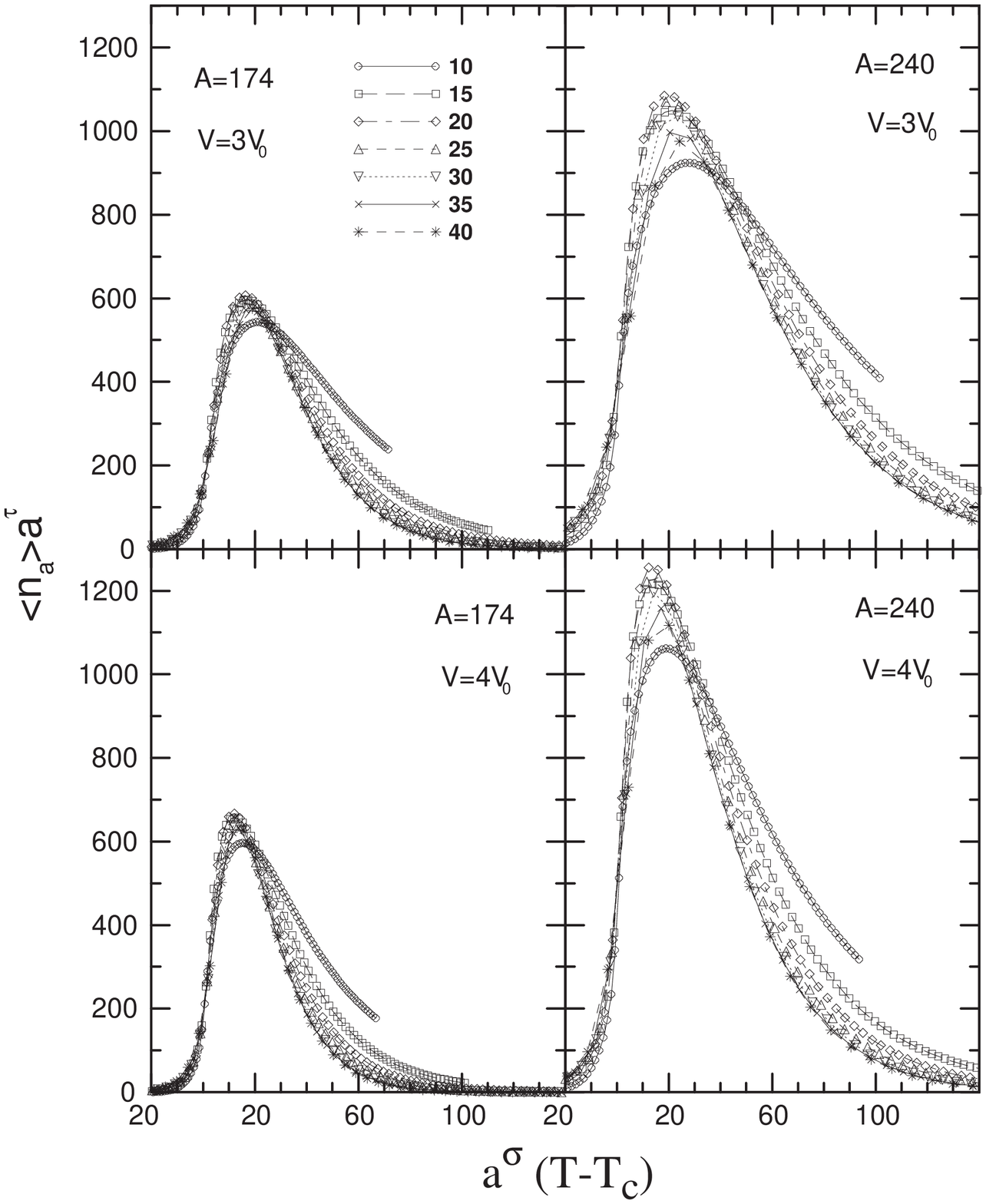}
\end{center}

\vskip 0.2in
Fig.3. The scaling behaviour in the mass range $10\leq a\leq 40$ in the
thermodynamic model for different systems at different freeze-out
densities.

\vskip 0.2in
In \cite{Scharenberg} the Copenhagen SMM is used to show that approximate
scaling is obtained.  The hope would then be that the theory also
demonstrated criticality.  The SMM is, in spirit, very close to the
thermodynamic model, thus we doubt that the very approximate collapse
of $\langle n_a \rangle a^{\tau}$ on one curve is any indication of criticality.
It is impossible to disentangle what errors
arise because the wrong formula is applied and what errors arise
because of the finiteness of the system and many other factors such
as the coulomb force, pre-equilibrium emission etc.
Experimental data would have a very hard
time of choosing between a first and a second order transition.

From 8 GeV/c $\pi^-$ on Au data, the ISiS \cite{Beaulieu} collaboration
obtained the caloric curve \cite{Ruangma}.  The specific heat was
obtained by differentiating with respect to $T$.  Experiment shows
that the peak of the specific heat coincides well with the position
where the $\chi^2$ for $\sum(\langle n_a \rangle - Ca^{-\tau})^2$ minimises.  Here
both $C$ and $\tau$ are taken as parameters to be fixed by minimisation.
The canonical model gives similar results.  Further details of
experiment and theory can be found in \cite{Das4} where effects
of the coulomb interaction on the position of the maximum of the specific
heat is discussed in detail.

We turn now briefly to another phenomenological model which was invoked
twenty years ago \cite{Goodman} but was revived recently \cite{Elliott3}.
This is yet another example where evidence for criticality can be
drawn too hastily.
Consider the formation of a droplet containing $a$ particles in the
liquid phase surrounded by $b$ particles in the gas phase.  At constant
temperature and pressure the Gibbs free energy is the relevant factor.
Then

$G_{with drop}=\mu_la+\mu_gb+4\pi R^2\sigma +T\tau$ ln$a$

and $G_{no drop}=\mu_g(a+b)$

The probability of formation of droplet containing $a$ particles is
proportional to $\exp(-\Delta G/T)$ so that the yield
of droplets of size $a$ is
\begin{eqnarray}
\langle n_a \rangle =Ca^{-\tau}\exp[(\mu_g-\mu_l)a/T+c_2a^{2/3}/T]
\end{eqnarray}
Here both $\mu_l$ and $\mu_g$ are functions of $T$.  At coexistence and
also at critical temperature, they become equal to one another.  Also
$c_2$ is a function of temperature and at $T_c$, the coefficient $c_2$
goes to zero.  Since above $T_c$, there is no distinction between the
liquid and the gas phase, one can not speak of droplets.  Thus the theory
only applies to $T < T_c$.  As such, the formulation is more limited
than that of eq.(15) which applies to both sides of $T_c$.  We now generate
values of $\langle n_a \rangle$ from the thermodynamic model for different temperatures
and try to fit these ``data'' using eq.(16).  The following fit was tried.
We set $\tau$=2.  Let $\alpha=(\mu_g-\mu_l)/T$, $\gamma=c_2/T$.  We fit
the calculated $\langle n_a \rangle$ to $Ca^{-2}\exp(\alpha a+\gamma a^{2/3})$ at
different temperatures where $\alpha,\gamma$ values at each temperature
are varied for best fit.  The values of $\alpha,\gamma$ as functions of
temperature are shown in fig. 4 where we also show that the parametrisation
fits the values of $\langle n_a \rangle$ very accurately.
The values of $\alpha$ and $\gamma$
both go to zero near temperature $T$=6.5 MeV suggesting the critical
temperature is 6.5 MeV.  Of course this conclusion would again be
wrong since
the model which gave these $\langle n_a \rangle$'s has only first order phase transition.

One problem is that whenever a fit, whether through eq.(15) or through
eq.(16), is done, the fit is attempted for a narrow range, $a$=6 to 40.
In this limited range moderate to excellent fits are obtainable for
different looking parametrisations.  It is shown in \cite{Das1} that if the
range of $a$ could be extended to beyond 100, different parametrisations
would diverge.  Unfortunately, the range of $a$ has to be limited.
For example higher values of $a$ would have contamination from
fission processes which is something we do not wish to include.
If we are stuck to a limited range of $a$'s we will also be limited
by ambiguity.

\vskip 0.2in
\begin{center}
\includegraphics[width=4.5in,height=4.5in]{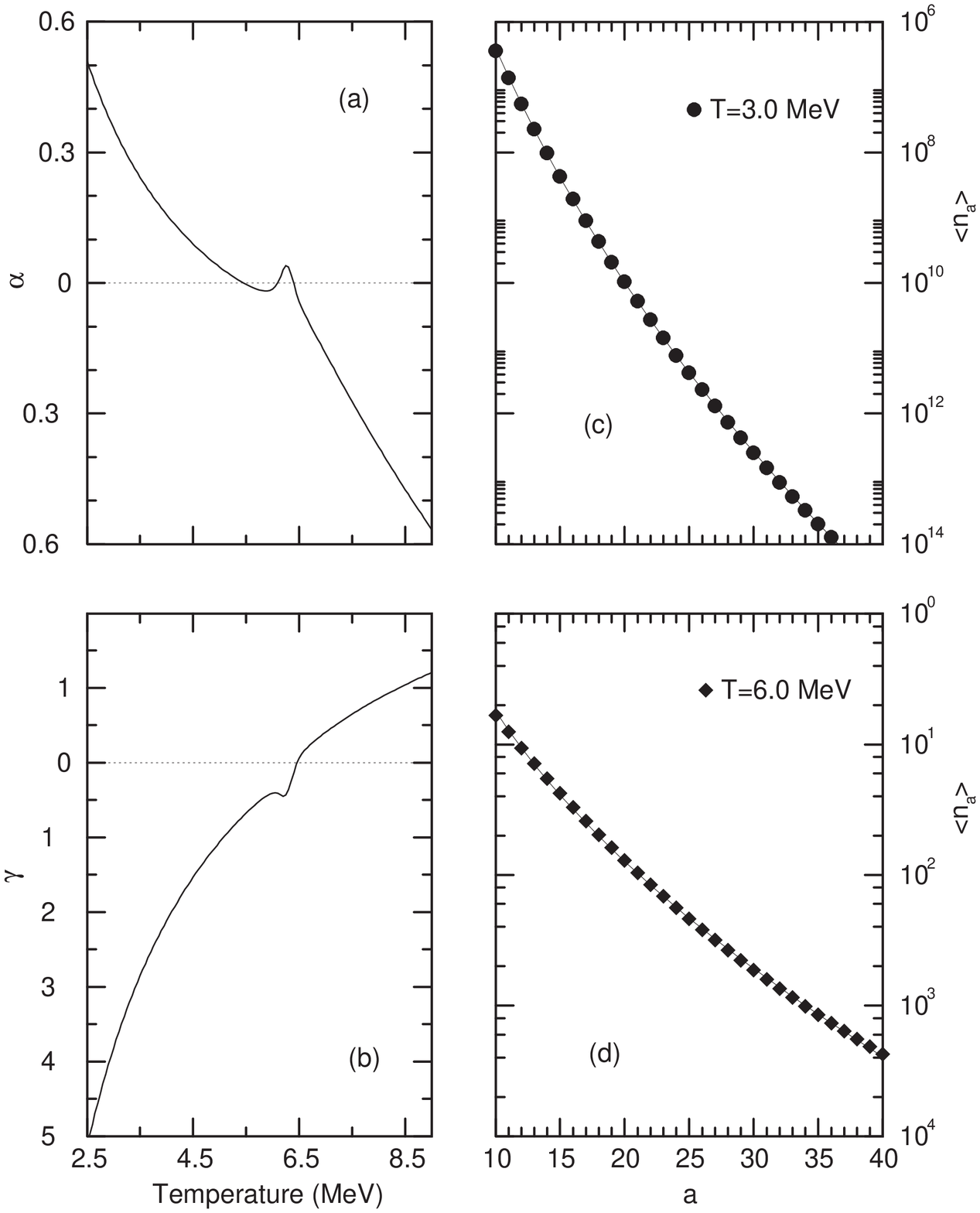}
\end{center}

\vskip 0.2in
Fig.4. The parameters of the droplet model $\alpha$ and $\gamma$ as a
function of temperature for a system of 240 particles at freeze-out 
volume $4V_0$.  The right panels show the fit of the model to the
yields obtained in the thermodynamic model

\vskip 0.2in
The emphasis towards unravelling critical phenomena from data on
intermediate energy heavy ion collisions is at least partly due
to history.  The observation by the Purdue group \cite{Finn} that
the yields of the fragments produced in $p+Xe$ and $p+Kr$ obeyed a
power law $\langle n_a \rangle \propto a^{-\tau}$ led to a conjecture that the
fragmenting target was near the critical point of liquid-gas phase
transition.  The origin of this conjecture is the Fisher model
\cite{Fisher} which predicts that at the critical point the yields of the
droplets will be given by a power law.  Also the first microscopic
model that was used \cite{Bauer1,Campi1}
to compute yields of fragments was the percolation
model which has only a continuous phase transition and a power law
at criticality.
The power law is no longer
taken as a ``proof'' of criticality.  There are many systems which
exhibit a power law: mass distribution of asteroids in the solar
system, debris from the crushing of basalt pellets \cite{Hufner}
and the fragmentation of frozen potatoes \cite{Oddershede}.  In
fact the lattice gas model which has been used a great deal for
multifragmentation in nuclei gives a power law at the critical point,
at the co-existence curve (this is a first order phase transition
provided the freeze-out density is less than half the normal
density) and also along a line in the $T-\rho$ plane away from the
coexistence curve \cite{Pan2,Chomaz1,Dasgupta5}.

We finish this section stating that the lattice gas model which
has a Hamiltonian and can be and has been used to fit many
data (not in any obvious way connected with phase transition)
also predicts a first order phase transition at intermediate
energy \cite{Pan2,Chomaz1}.

\subsection{Comparision with mean-field theory}

Here we concentrate on the thermodynamic model but as applied to
nuclei with neutrons and protons.  The operative equations are (10)
to (13) but we will switch off the coulomb term ($\kappa$
of eq.13 will be set to zero).  The objective is to compare with
finite temperature Hartee-Fock results for nuclear matter.  For
nuclear matter the coulomb interaction has to be switched off and
one retains only the nuclear part of the interaction.

Phase transitions are often considered in the mean-field model.  Examples
for the present discussion are
\cite{Jaqamann,Bertsch1,Dasgupta3}.  Invariably a grand canonical ensemble
is used characterised by a neutron chemical potential $\mu_N(T,\rho)$
and a proton chemical potential $\mu_P(T,\rho)$.  The use of the
grand canonical model would imply that the results are valid for very
large systems although in nuclear physics we often use the grand
canonical ensemble for not so large systems as well.

Muller and Serot \cite{Muller1,Muller2} used the mean-field model
to investigate phase transition in nuclear matter.  Normally nuclear
matter means a very large system with $N=Z$ with the coulomb force
switched off.  For this section we will use
the term nuclear matter for very
large systems but $N$ can be different from $Z$.  The coulomb is
switched off as usual.  Define proton fraction $y=Z/(N+Z)$.  Symmetric nuclear
matter has $y$=0.5 and would have a first order phase transition
below the critical point.  But for $y$ deviating significantly
from 0.5, these authors demonstrate with a more general
Maxwell like construction that
the first order phase transition would turn into second
order.  Further the phase transition would take place neither
at constant volume nor at constant pressure but would have a more
general path to traverse.

The general characteristics of mean-field theories is that one
is constrained to have one density.  Having the same density everywhere is a
big price to pay.  For example, this would not permit a
liquid phase at one place and a gas phase at another.  The limitation
of one density only shows up as mechanical instability, i.e.,
in parts of the equation of state diagram ($p-\rho$ isothermals)
$\partial p/\partial \rho$ turns out to be negative.  This is
unacceptable for infinite matter
and then one has to, by hand, correct
this using a Maxwell construction.  The thermdynamic model is very
different.  Here, for example, $\rho/\rho_0=0.3$ does not mean
that at the freeze-out volume, matter is uniformly stretched.  Rather
matter breaks up into different blobs all with the same density $\rho_0$
but there are empty spaces between blobs.  If there is a large blob,
we identify it as liquid, nucleons and light composites in the adjoining
spaces form the gas (In \cite{Bhat2}, it is shown that this last
scenario has a lower free energy compared to uniform stretching as
asumed in Hartree-Fock theory).  For large matter, there is no need for
$(\partial p/\partial\rho)_T$ to be negative.

A similar thing happens with isospin fractionation.  In mean-field theory,
there is one value of $y$ everywhere.  Experimentally, it is verified
that if the dissociating system has a small $y$, then after break
up, the largest blob has $y > y_{diss}$ whereas $n_p/(n_p+n_n) < y_{diss}$
where $n_p, n_n$ are free protons and free neutrons respectively.
Here $y_{diss}$ is the the $y$ value of the dissociating system.
One might say the liquid phase has a different $y$ from that of the
gas phase.
Again mean-field theory would have a hard time accommodating this.
It must have the same value of $y$ everywhere and the fact that
this is an unstable situation shows up in the following way.
If we draw $\mu_P(\mu_N)$ as a function of $y$ at constant temperature,
the derivative
$(\partial \mu_P/\partial y)_p$ can turn out to be negative (equivalently
$(\partial\mu_N/\partial y)_p$ can turn out to be positive).
In the thermodynamic model, isospin fractionation happens naturally.
In general, the model has, as final products, all allowed composites,
$a,b,c,d$...., where the composite $a$ has $y_a=i_a/(i_a+j_a)$ where
$i_a,j_a$ are the proton and neutron numbers of the composite $a$.
The only law of conservation is $Z=\sum_ai_a\times n_a$ and
$N=\sum_aj_a\times n_a$.  So a large chunk can exist with higher $y$
than that of the whole system and populations of other species can
adjust to obey overall conservation laws.  Whatever partition lowers the
free energy will happen.  Since we are using a canonical model, we
do not need the chemical potentials $\mu_P$ or $\mu_N$ but we can
compute them anyway from the relation $\mu=(\partial F/\partial n)_{V,T}$.
We know the values of $Q_{Z,N}, Q_{Z-1,N}$ and $Q_{Z,N-1}$.  Since
$F=-T$ln$Q$, one has $\mu_P=-T$(ln$Q_{Z,N}$-ln$Q_{Z-1,N}$) and
similarly for $\mu_N$.

Calculations with the canonical model discussed in this article,
do not show regions of
negative $(\partial\mu_P/\partial y)_{p,T}$ \cite {Das2,Das3}.
These also suggest that the phase transition in the canonical
model remains first order for asymmetric matter.  We show in fig.5
results of $c_V$ calculation for different degrees of asymmetry.
One sees that as the system gets bigger, the maximum in $c_V$
becomes narrower and higher, ensuring there will be a break
in the first derivative of the free energy in the large matter
limit.

Chemical instability for finite systems in Hartee-Fock theory
has also been worked out.  Contributions of both coulomb and
surface terms can be included.  For details see
\cite{Lee1,Lee2,Lee3}.

\vskip 0.2in
\begin{center}
\includegraphics[width=3.5in,height=3.5in]{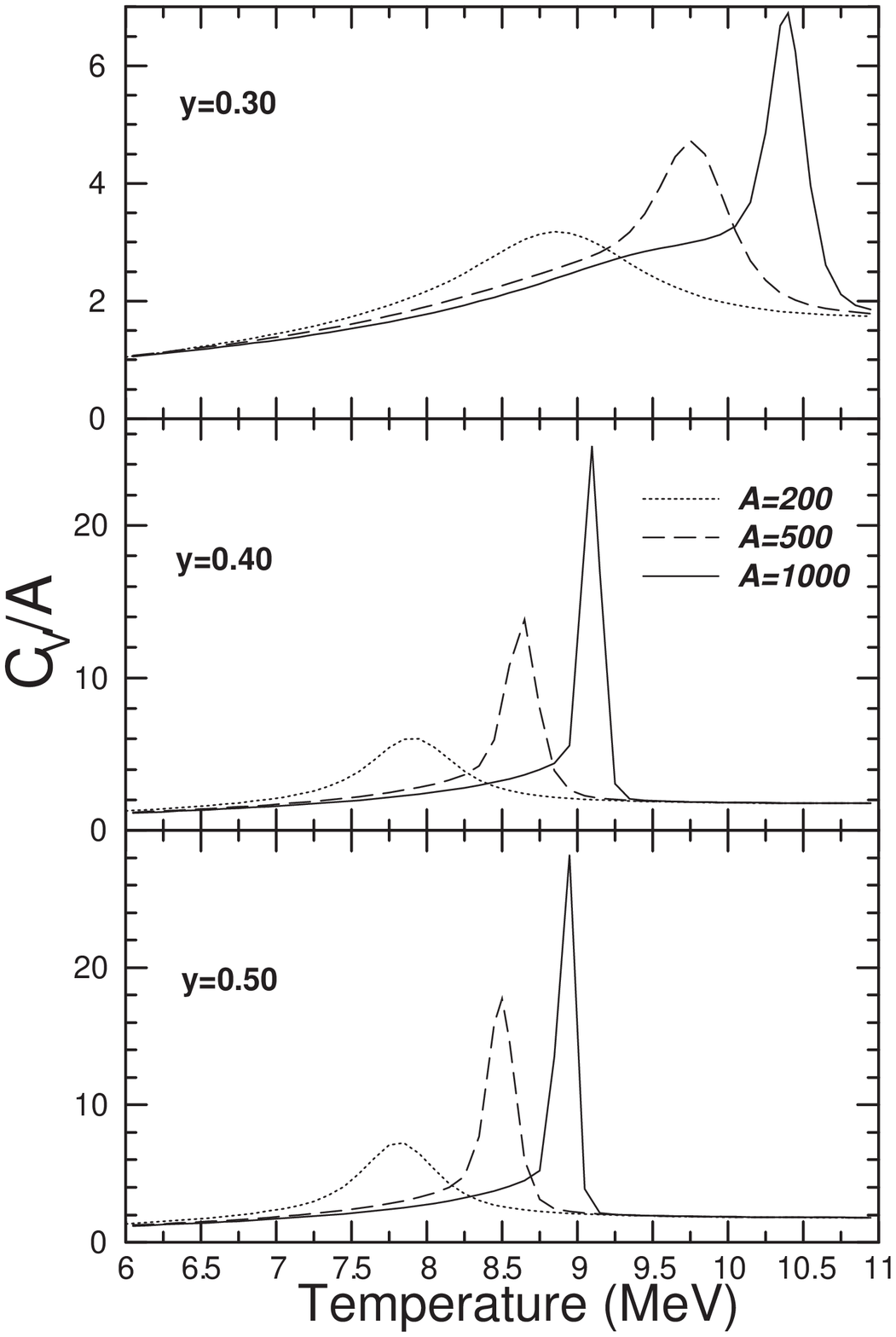}
\end{center}

\vskip 0.2in
Fig.5. The $C_V/A$ as a function of temperature for systems of
200,500 and 1000 particles with different proton fractions:
($y=Z/A$).

\vskip 0.2in
\section{Comparison of canonical and grand canonical}
As noted in the introduction, the grand canonical version of
the model we are pursuing in this paper has been known and used for a
long time.  Now that we know how to treat an exact number of particles
rather than an ensemble of particle numbers, it will be useful in a
few cases to examine, given that our dissociating system has an exact
number of particles, how the use of grand canonical ensemble affects
the prediction of observables.  For simplicity, we start with the model
of one kind of particles and our dissociating system has 200 particles.
Thus we can have monomers, dimers, trimers,...upto a composite of 200
particles.  In the grand canonical ensemble, the average number of
composites with $k$ nucleons is given by
\begin{eqnarray}
\langle n_k \rangle =\exp(\beta\mu k)\omega_k=\exp(\beta\mu k)V\tilde\omega_k
\end{eqnarray}
Here $\beta$ is the inverse of temperature and $\omega_k$ is the same as
defined in eq.6 and $\mu$ is the chemical potential.  We also use
$\tilde\omega=\omega/V$ where $\tilde\omega$ only depends upon the
composite and the temperature but not upon the volume of the dissociating
system.  The chemical potential is determined by solving
\begin{eqnarray}
\rho=\sum_{k=1}^{k_m}k\exp(k\beta\mu)\tilde\omega_k
\end{eqnarray}
In this example $k_m$=the number of particles in the largest cluster=200.
Having determined $\mu$ we now
find $\langle n_k \rangle$ from $\langle n_k \rangle =\exp(k\beta\mu)V\tilde\omega_k$.

\vskip 0.2in
\begin{center}
\includegraphics[width=3.5in,height=3.5in,angle=-90]{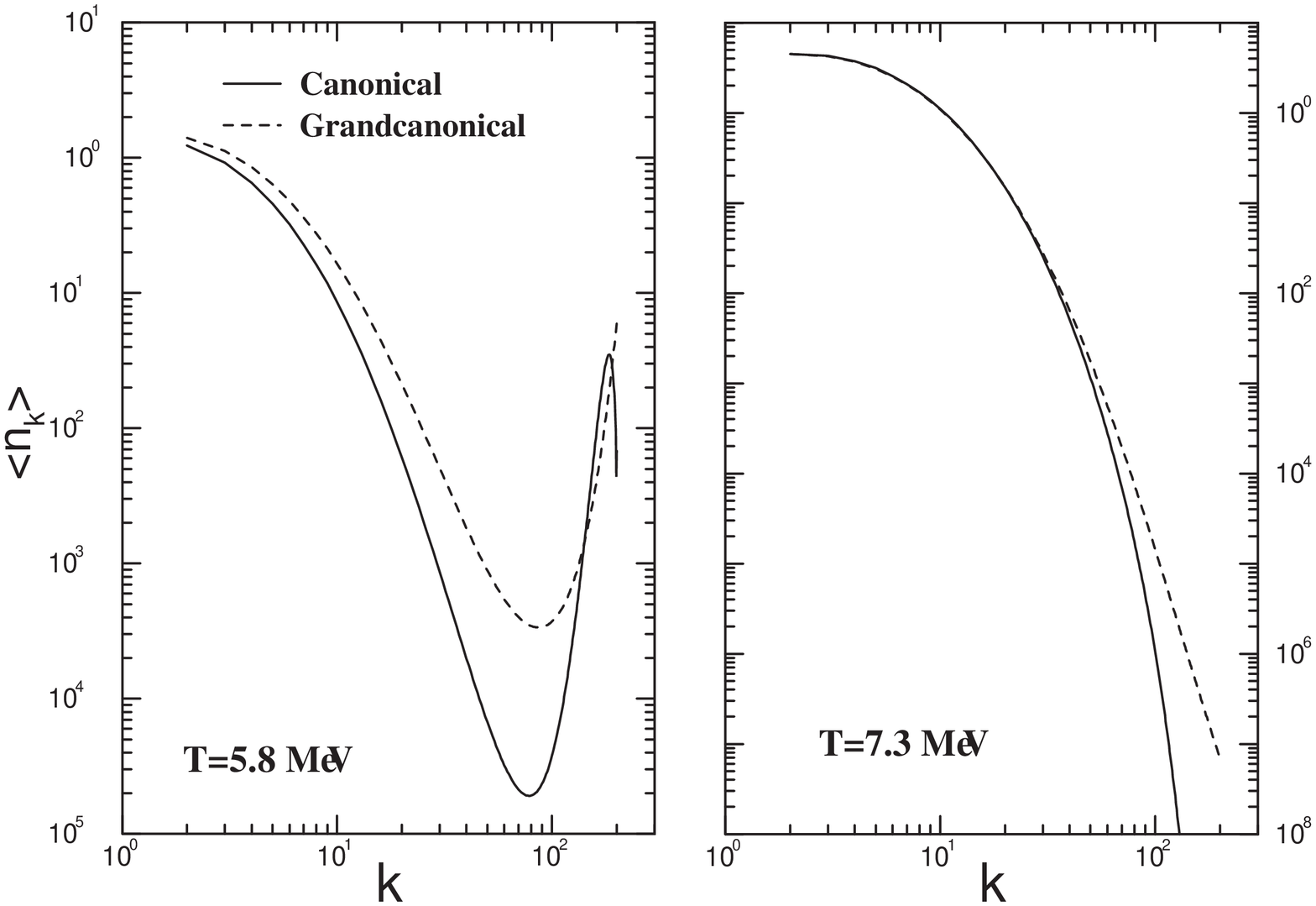}
\end{center}

\vskip 0.2in
Fig.6. Comparison of yields obtained in the canonical and grand canonical
models at different temperatures, for a system of 200 particles
at freeze-out volume $3V_0$.

\vskip 0.2in
In figs. 6 and 7, we make a comparison of $\langle n_k \rangle$'s from 
canonical and
grand canonical ensembles.  The value of $V$ was set at $3V_0$.
Results are shown for temperatures below the phase transition
temperature and above it.  Fig. 6 seems very reasonable.  The overall features
are similar.  The differences get highlighted in fig. 7.  At temperature
7.3 MeV,
$\langle n_k \rangle^{GC}$ and $\langle n_k \rangle^{C}$ are practically the same upto $k$=40 but
deviate wildly afterwards.  Since most of the time we are not interested
in the heavier products and $k$=40 is the limit of 
intermediate mass fragments one is investigating,
the grand canonical ensemble does an adequate job.  We must be aware
however, that, for heavy composites the grand canonical ensemble does a
very poor job.  The accuracy of the grand canonical ensemble at
temperature 5.8 MeV (below the phase transition temperature) is absolutely
awful for almost all composites.  This is also the temperature
range appropriate for most intermediate energy reactions.  It is thus
dangerous to use the grand canonical ensemble in intermediate
energy heavy ion reactions.

\vskip 0.2in
\begin{center}
\includegraphics[width=2.5in,height=2.5in]{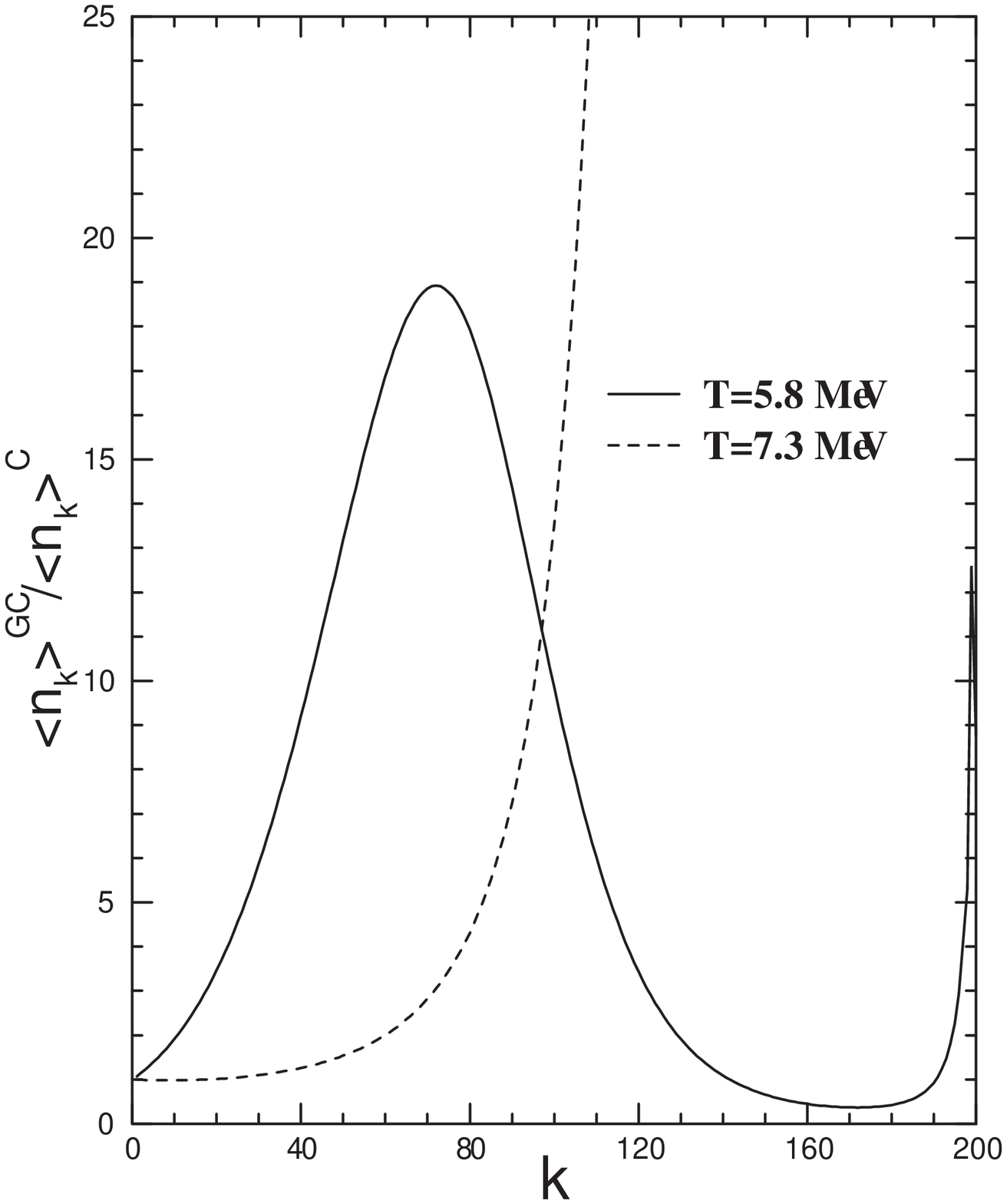}
\end{center}

\vskip 0.2in
Fig.7 The ratio of yields obtained in the grandcanonical and the
canonical model at different temperatures.

\vskip 0.1in
If however, one is only interested in finding the ratio of
populations of two adjacent composites, the grand canonical continues
to be useful over a larger domain.  This is shown in fig.8.
\vskip 0.2in
\begin{center}
\includegraphics[width=2.5in,height=2.5in,angle=-90]{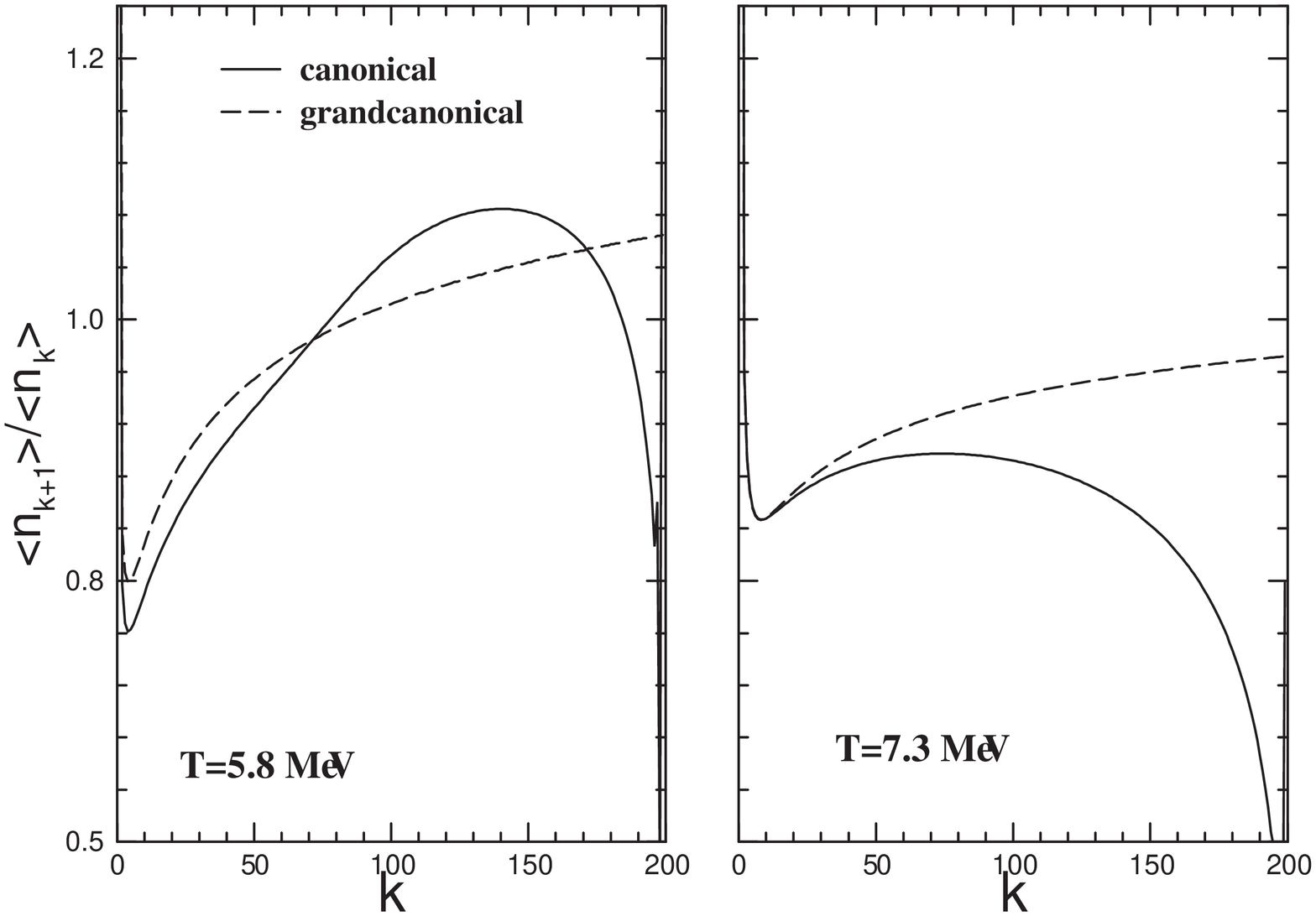}
\end{center}

\vskip 0.1in
Fig.8 Ratio of yields of adjacent composites in the two models.

The very different populations of composites below the phase transition
temperature leads to drastically different caloric curves in the grand
canonical ensemble and canonical ensemble.
As noted in section 4 and shown in fig.2, for a fixed density
the specific heat per particle maximises at a certain temperature.
Keeping density fixed, if we increase the number of particles the
height of the maximum increases and the width decreases.  In fig.9
we show this again for 200 and 2000 particles, but now we have also
indicated the specific heat calculated in the grand canonical
ensemble.  In both the models, the peak of the specific heat increases
when we go from 200 to 2000 particles and the widths decrease but the
results are much more dramatic in the canonical model.  In particular,
it is not obvious that the specific heat in the grand canonical ensemble
will attain extraordinary heights and miniscule widths.  In fact, it was
suggested in the literature, engaging the grand canonical ensemble, that
there is a discontinuity in the value of the specific heat at phase
transition but no infinity \cite{Bugaev}.
\vskip 0.2in
\begin{center}
\includegraphics[width=3.5in,height=3.5in,angle=-90]{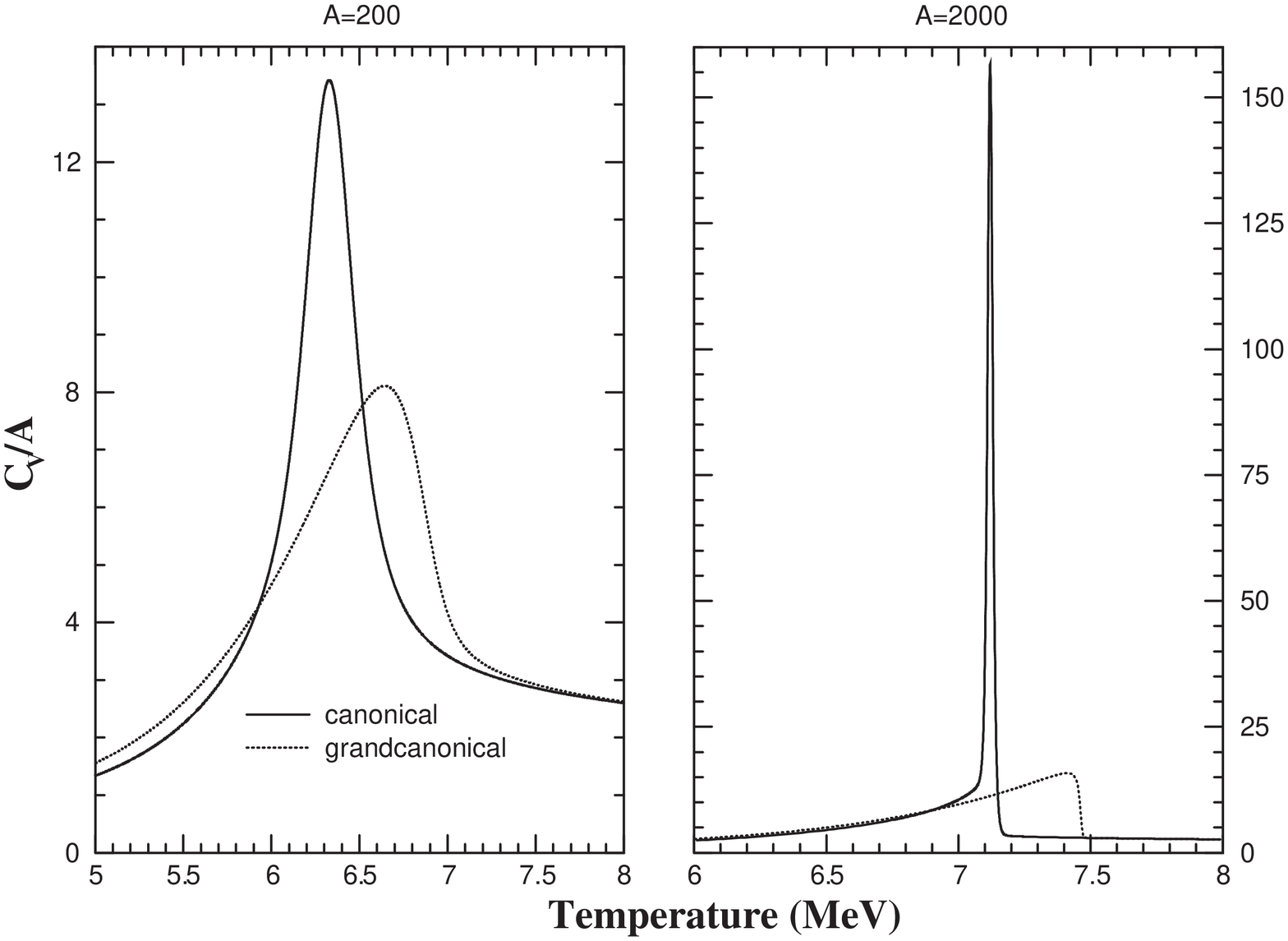}
\end{center}

\vskip 0.2in
Fig.9 Specific heat per particle at constant volume when the system
has total number of particles 200 and 2000.  Canonical and grand
canonical values are shown.

\vskip 0.1in
To understand at a more fundamental level the cause of the difference
in values of specific heats in the two ensembles,
we will analyse the case of 2000 particles in some detail.  In the grand
canonical model, even though we are using the average value of the
particle number to be 2000, there are, in practice, systems with varying
particle numbers (in principle, from 0 to $\infty$).  The part which
has, for example, 1000 particles has density half of the prescribed
density.  The peak in the specific heat of this half density will occur
at a different temperature than that which maximises the specific
heat at density 2000/V.  Thus there is a smearing effect.
This is always an inherent problem with
using the grand canonical ensemble but most of the time the fluctuation
from the average value is small enough that one can live with it.
This would have meant, in our present example, the part which contains
1000 particles is so negligibly small that it does not matter.  This
however is not so in the present model below the phase transition
temperature.

In the present case, the grand canonical calculation starts out by
obtaining $\mu$ from eq.18 where $k_m$=2000; $\rho$ was taken
to be $\rho_0/2.7$.  The average value of $\langle n_k \rangle$
is then given by
eq.(17) where $V=2000\times 2.7/\rho_0$.  With this we have
$\sum_{k=1}^{k=2000}k \langle n_k \rangle =2000$.
The fluctuations in the model can be calculated easily.  We have the
general statistical relation
\begin{eqnarray}
\frac{1}{\beta^2}\frac{\partial^2lnQ_{gr.can}}{\partial^2\mu}=
\langle N^2 \rangle - \langle N \rangle^2
\end{eqnarray}
Here $Q_{gr.can}$ is the grand canonical partition function.  We can
write two expressions for $Q_{gr.can}$.  One is:
\begin{eqnarray}
lnQ_{gr.can}=\sum_{k=1}^{k=2000}\exp(\beta\mu k)\omega_k
\end{eqnarray}
This immediately leads to
\begin{eqnarray}
\langle N^2 \rangle - \langle N \rangle^2 =\sum_{k=1}^{k=2000} k^2 \langle n_k \rangle
\end{eqnarray}
which is easily calculable.
The other expression we can exploit in the present case is
\begin{eqnarray}
Q_{gr.can}=\sum_{K=1}^{\infty}\exp(\beta\mu K)Q_{K,k_m}
\end{eqnarray}
where $Q_{K,k_m}$ is the canonical partition function of $K$
nucleons but with the restriction that the largest cluster can not
have more than $k_m(=2000)$ nucleons.  We can calculate these
explicitly using methods of section 2.  For practical reasons,
$K$ has to be cut off at the upper end.  Here we used $K$=10000
as the upper limit.  Since the average number of particles is 2000,
this appears to be a safe upper limit in eq. 22.  The quantity $\mu$ is
known from solving eq. 18.

The fluctuations calculated with eqs. 21 and 22 are shown in fig. 10.
One sees there is a temperature above which the fluctuations are
small.  At these temperatures, the grand canonical value of specific
heat is indistinguishable from the canonical value.  But as the
temperature is lowered, fluctuations grow rapidly and the results
begin to diverge.
\vskip 0.2in
\begin{center}
\includegraphics[width=3.5in,height=3.5in]{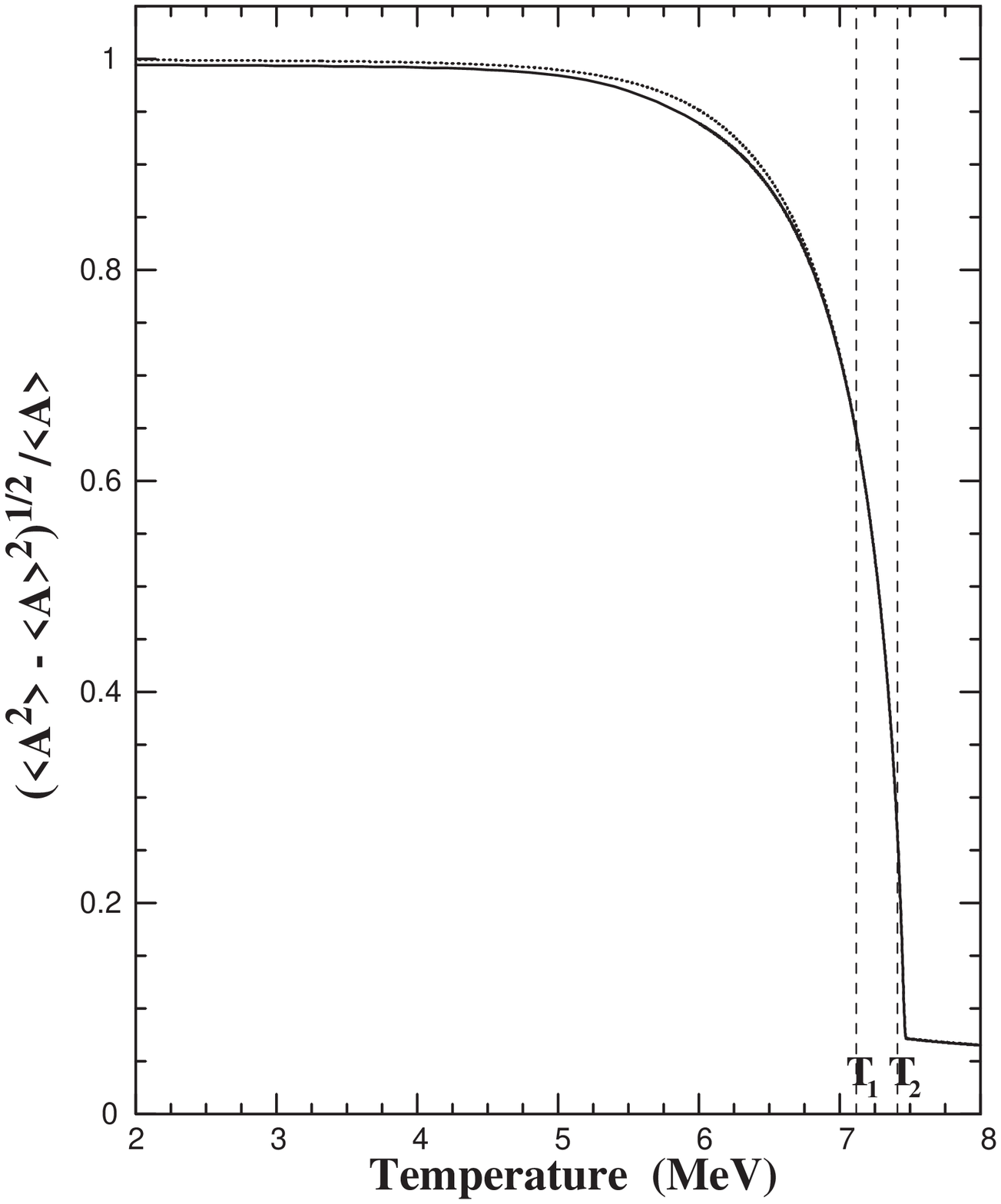}
\end{center}

\vskip 0.2in
Fig.10 Fluctuations calculated using eqs. (21) and (22).  The solid
line corresponds to using eq. (22) with $K$ cut off at 10,000 and
the dotted line corresponds to using eq. (21).  $T_1$ corresponds to
to the temperature where the specific heat maximises in the canonical 
calculation and $T_2$ to the temperature of highest specific heat
in the grand canonical calculation.

\vskip 0.1in

It is interesting to study fluctuations further.  The probability
of $K$ particles being in the grand canonical ensemble is
$\propto e^{K\beta\mu+ln Q_K}$.  We plot in fig.11 $\exp[\beta\mu(K-A)
+lnQ_K-lnQ_A]$.  This takes the value 1 at $K=A$ and in the normal
picture of the grand canonical ensemble would drop off rapidly on
either side of $A$.  This does happen at temperatures higher than the
boiling temperature.  The case at $T$=7.7 MeV corresponds to a
standard scenario.  But the situation at temperature 7.3 MeV is
drastically different.  The probability does not maximise at $K=A$
but at a lower value.  It is also very spread out with a periodic
structure.  The periodicity is 2000 and is linked with the fact
that in the case studied, the largest composite has 2000 nucleons
and at low temperatures, this composite will play a significant role.

More discussions on this case can be found in \cite{Das5}.

\begin{center}
\includegraphics[width=3.5in,height=3.5in,angle=-90]{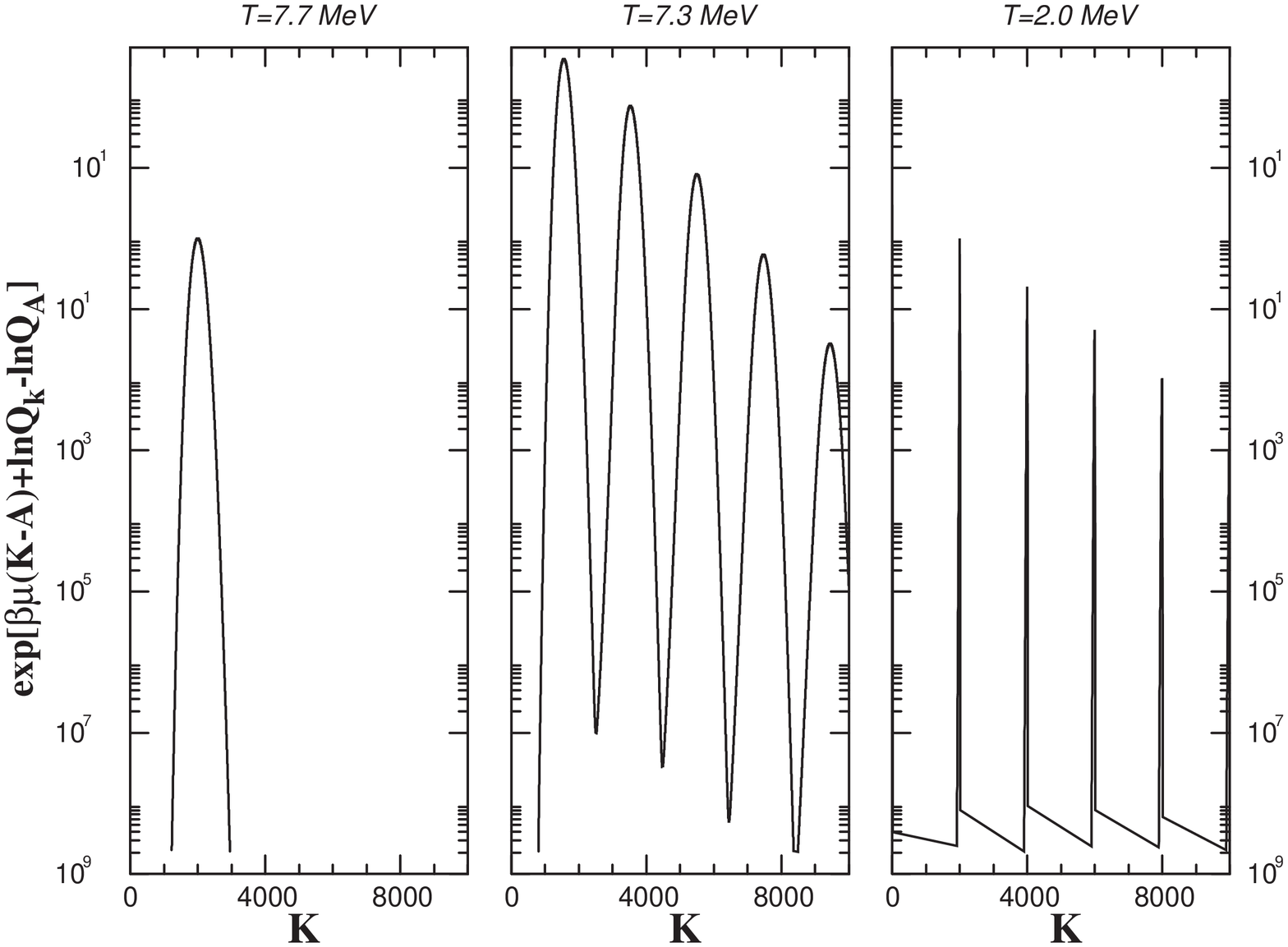}
\end{center}

\vskip 0.2in
Fig. 11.  These graphs show the spread of particle numbers in the 
grand canonical ensemble when the average particle number is 2000.
The spread is very narrow at temperature 7.7 MeV but becomes quite 
large at lower temperatures.

\vskip 0.1in
\section{Specific heat at constant pressure}
We have used $C_V$, the specific heat at constant volume a great deal
in the previous sections.  In canonical models $C_V$ is always
positive.  Writing: $\langle E \rangle =\frac{\sum E_i(V)\exp(-\beta E_i(V))}{\sum\exp(
(-\beta E_i(V))}$ and $C_V=(\partial \langle E \rangle /\partial T)_V$, we get
$C_V=\frac{1}{T^2} \langle ( E- \langle E \rangle )^2 \rangle$
which is the expectation value of
a positive definite operator.  However, specific heat at constant
pressure allows no such generalisations.  Here we enter into a discussion
of the specific heat at constant pressure in the thermodynamic model.
We should add that dissociation after two heavy ions collide is
largely an uncontrollable situation and we do not know what is a
better description: disassembly at constant volume, disassembly at
constant pressure or a hybrid situation.

Lately, interest in the topic has increased with the realisation
that for finite systems, $C_p$ can sometimes be negative and
such cases might arise in heavy ion collisions
\cite{Gross2,Chomaz2,Moretto2}.
To study this
possibility in our model, we find it convenient to look at the $p-\rho$
diagram at constant temperatures (isothermals).  This is shown in
fig.12. We see there are regions of mechanical instability where
$(\frac{\partial p}{\partial\rho})_T < 0$.  We will show that the
occurrence of negative $C_p$ happens in this region.

\vskip 0.2in
\begin{center}
\includegraphics[width=3.5in,height=3.5in]{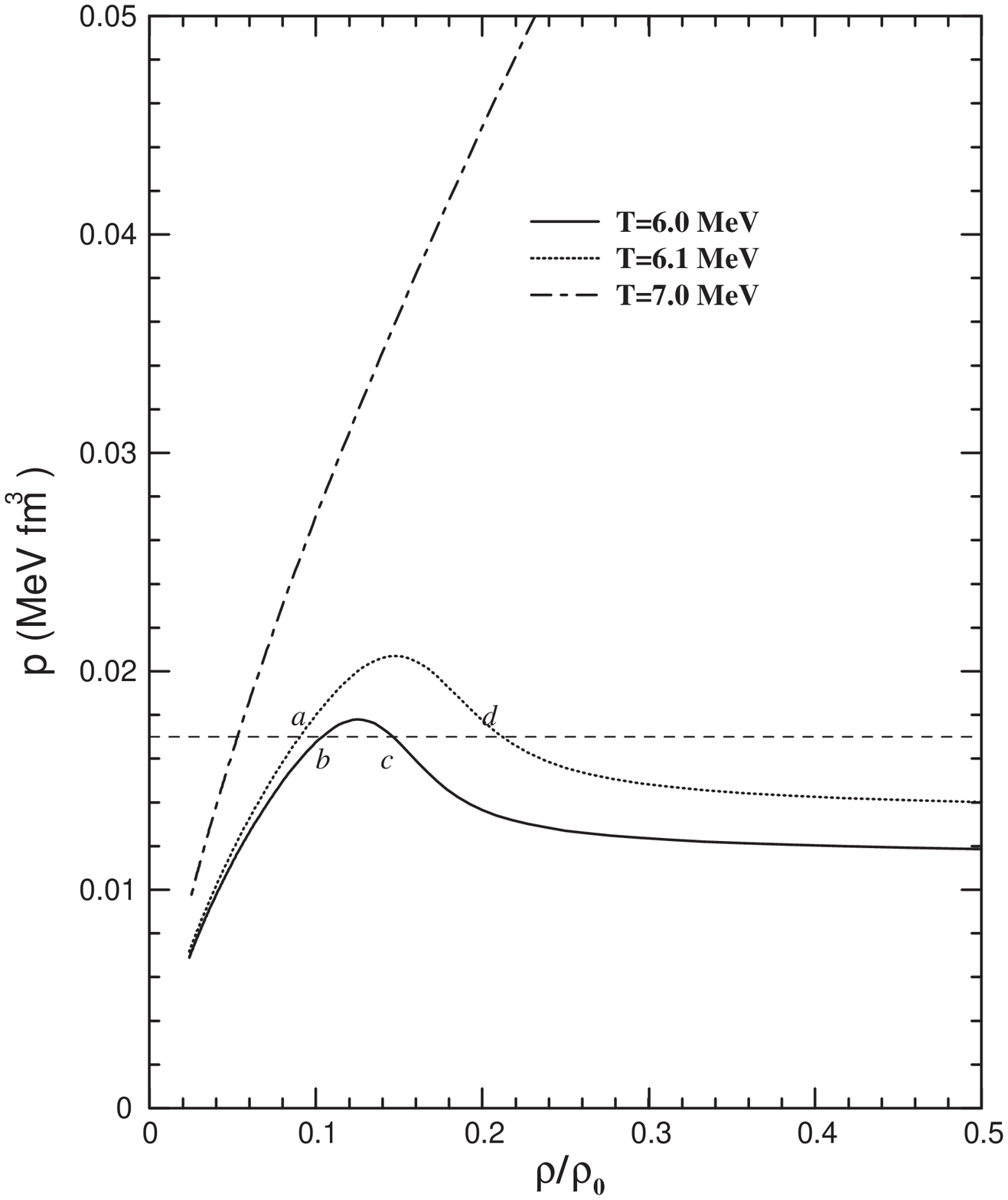}
\end{center}

\vskip 0.2in
Fig. 12.  EOS in the canonical model for a system of $A$=200
particles.  The largest cluster also has $N$=200.

\vskip 0.1in

The most famous case of mechanical instability is the Van der Waals
equation of state.  In nuclear physics, if one uses the Hartree-Fock
theory, then also large regions of mechanical
instability appear.  Examples of this can be seen in many published
works: \cite{Jaqamann,Dasgupta2,Das2}.  All these published works
are for infinite systems (unlike the $p-\rho$ diagram for fig.12
which is drawn for 200 particles).  Quantitative examination of the
equation of state diagrams reveal that the regions of mechanical instability
are far bigger in the case of Hartree-Fock as opposed to what
we see in fig.12. In fact, plotted on the same scale, the region of
mechanical instability
would be tiny (ref. fig.1 of \cite{Das2}) and one would have
to plot it in an expanded scale (such as is done in fig.12) to
study it quantitatively.

In the Van der Waals case or in the Hartree-Fock case for infinite 
nuclear matter
one uses a Maxwell construction to replace the region of mechanical
instability \cite{Reif}.  In the thermodynamic limit, regions of mechanical
instability should disappear.  In our case there is no prescription
for Maxwell construction.  Also since our system is very finite, we take
the mechanical instability in fig.12 as real and follow the consequences
for the specific heat.  In the figure we have drawn isothermals at
three temperatures; $T_1 < T_2 < T_3$.  Here $T_2$ is only slightly higher
than $T_1$.  Instead of $\rho$ let us use the variable $V\propto 1/\rho$.
The pressure is given by $p=T(m/V)$ where $m$ is the multiplicity.
[We actually use $m-1$ but this is inconsequential for the discussion
to follow.] For
the simple case of monomers only, $p$ is given by $p=T(A/V)$ where $A$
is the number of particles.  This number does not change thus $p$
keeps falling with $V$.  In our case, $m$
is significantly less than $A$.  It is not a constant as $V$ and/or $T$
change.  As can be readily guessed, $m$ increases with $T$ at constant
$V$; $m$ also increases with $V$ at constant $T$.  Negative compressibility
is marked by $(\partial m/\partial V)_T > m/V$.

Let us consider the points $c$ and $d$ in fig. 12.  Let $c$ have multiplicity
$m$, volume $V$ and temperature $T$; for $d$ the corresponding quantities
are $m+\delta m, V+\delta V$ and $T+\delta T$.  Here $\delta V$ is
negative, $\delta T$ is positive.  Using
\begin{eqnarray}
p=T\frac{m}{V}=(T+\delta T)\frac{m+\delta m}{V+\delta V}
\end{eqnarray}
we arrive at
\begin{eqnarray}
\frac{\delta m}{m}=\frac{\delta V}{V}-\frac{\delta T}{T}
\end{eqnarray}
In the region ($c,d$), $\delta V$ is negative, $\delta T$ is positive thus
$\delta m$ is negative.  If $m$ goes down then so does the potential
energy (creating more $m$ creates more surface and hence increases energy).
The change in kinetic energy is :
$\frac{3}{2}[(m+\delta m)(T+\delta T)-mT]$ which using eq.24 is
$\approx\frac{3}{2}\frac{\delta V}{V}mT$.  This is negative also.  Thus
both kinetic and potential energy fall giving rise to negative $C_p$.  If
on the other hand we consider points $a$ and $b$, point $a$ has both
a bigger volume and a bigger temperature thus $\delta m$ is positive.
This would make both the kinetic and potential energy rise when one moves
from $b$ to $a$.  This is illustrated in Table 1. The caloric curve
of fig. 13 shows regions of negative $C_p$.

\vskip 0.1in
\begin{center}
\includegraphics[width=3.5in,height=3.5in,angle=-90]{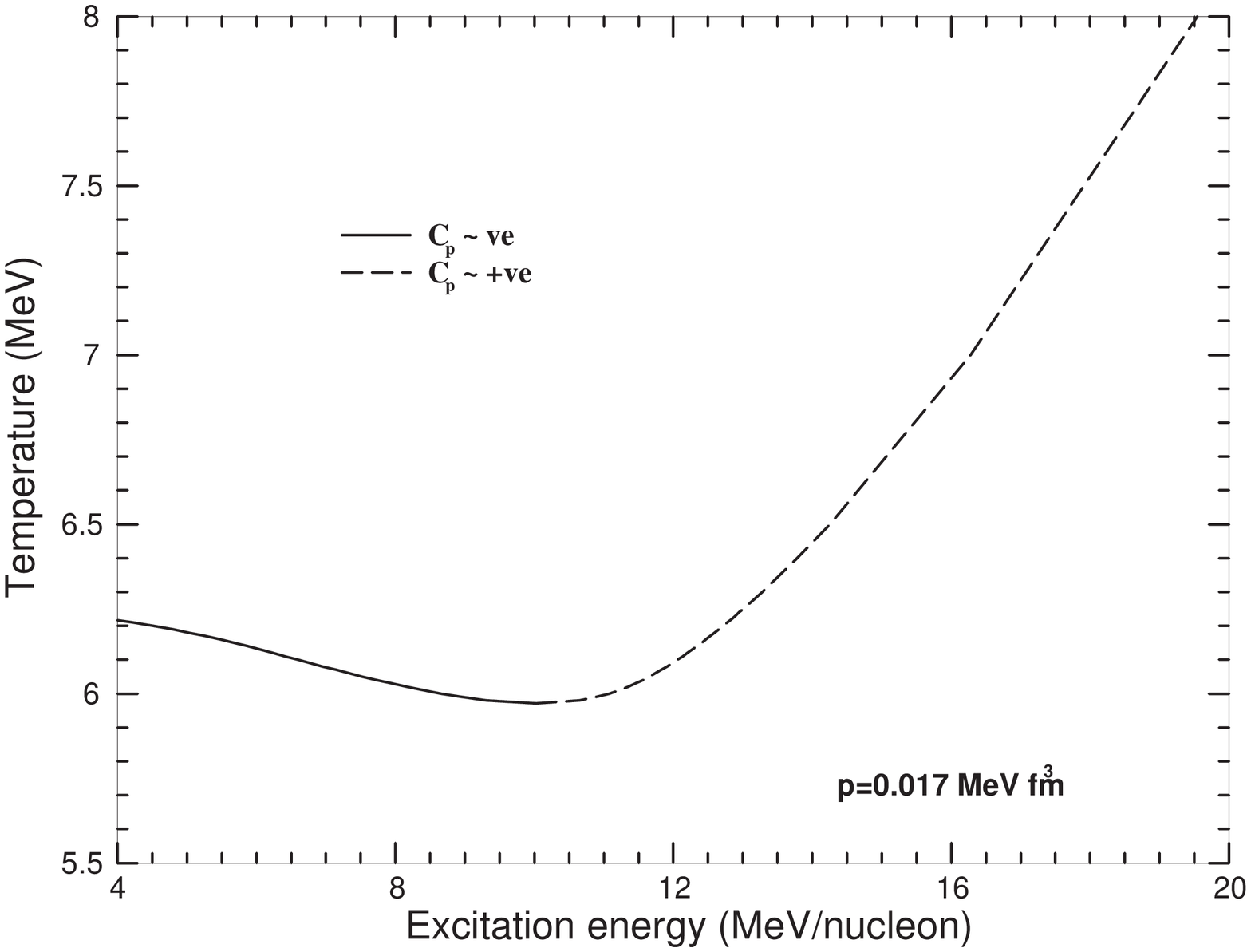}
\end{center}

\vskip 0.1in
Fig. 13.  Caloric curve at constant pressure ($p$=0.017 MeV$fm^{-3}$)
in the canonical model with$A$=200 and $N$=200.  The solid and dashed
portions of the curve give -ve and +ve $C_p$ respectively.

\vskip 0.1in
\begin{table}
\begin{center}
\caption{ \it Variation of energies per particle $(MeV)$ with
temperature $(MeV)$ in the negative and positive compressibility
zones, for $p = 0.017 \ MeV \ fm^{-3}$}
\begin{tabular}{cccccc}
\hline
\multicolumn{1}{c}{} &
\multicolumn{1}{c}{$T$} &
\multicolumn{1}{c}{$\rho/\rho_0$} &
\multicolumn{1}{c}{$e_k/A$} &
\multicolumn{1}{c}{$e_{pot}/A$} &
\multicolumn{1}{c}{$e_{tot}/A$} \\
\hline
&6.0&0.146&0.978&-5.235&-4.257 \\
$\frac{\partial p}{\partial \rho} < 0$&6.1&0.212&0.638&-6.970&-6.332 \\
&6.2&0.392&0.294&-8.708&-8.414 \\
\hline
&6.0&0.104&1.422&-3.271&-1.849 \\
$\frac{\partial p}{\partial \rho} > 0$&6.1&0.090&1.653&-2.513&-0.859 \\
&6.2&0.082&1.824&-2.027&-0.202 \\
\hline
\end{tabular}
\end{center}
\end{table}

Let us consider the thermodynamic limit.  This will be reached when
the number of composites near the boundaries of the freeze-out
volume is negligible to the number of composites well inside.  In this
limit, intensive variables remain unchanged when extensive variables
are changed by a constant factor.  Thus if $A$ is the total number of
nucleons in the system and we change, at constant temperature,
$A\rightarrow A+\alpha A, V\rightarrow V+\alpha V$ the pressure
$p=\frac{m}{V}T$ must remain constant.  This means, for constant $T$,
when $A\rightarrow A+\alpha A, V\rightarrow V+\alpha V$, $m$ must
change to $m\rightarrow m+\alpha m$.  Now for compressibility, $A$
stays at $A$, but $V$ to $V+\alpha V$ thus $m$ must change to less than
$m+\alpha m$.  Then the pressure will fall when $V$ is increased, i.e.,
regions of negative compressibility disappear.

It would be nice to demonstrate this feature directly by doing
canonical calculations for larger and larger systems.  The area
over which negative compressibility appears does drop
as larger and larger systems are used but the
convergence is slow.  Instead we will use the grand canonical ensemble
to get to the $A=\infty$ limit.  For a given density we solve eq.18,
setting once $k_m=200$ and $k_m=2000$, the other time.  This means, in the
first case, the largest composite has 200 nucleons and in the second
case, the largest composite has 2000 nucleons.  The temperature
is chosen to be 6 MeV.  Eq.18 has no reference to either $A$ or $V$
(only their ratio),
the implication being for the grand canonical model to be good each factor
is $\infty$ or very large.
Pressure in the grand canonical ensemble is
given by $p=(T/V)lnQ_{grand}$ which leads to
$p=\sum_{k=1}^{k_m}\exp(k\mu\beta)\tilde\omega_k$.

Fig. 14 compares canonical calculation with $A$=200 and $k_m\equiv N=200$
with $A=\infty$ and $N$=200.  We see in the low density (the gas phase)
the two diagrams coincide.  The rise of pressure with density is quite
rapid and linear.  After the two diagrams separate, the rise of pressure
with density in the grand canonical ensemble slows down considerably but
there is no region of mechanical instability although the canonical
calculation with 200 particles has a region of instability.  In the
grand canonical result which represents the thermodynamic extrapolation,
we have not reached the classic liquid-gas coexistence limit where there
would not be any rise of pressure at all (such as in Maxwell's construction).
We think the reason is this.The largest cluster is 200 which is not a big
enough number.  We now increase the largest cluster size to 2000.  Now the
coexistence region is very clear and there is unmistakable signature
of first order phase transition.  In the same figure we also show results
of canonical calculation with $A$=2000 and $N$=2000.  The region of
mechanical instability has gone down considerably but it has not disappeared
showing that we have not reached the thermodynamic limit yet.

Thermodynamics allows $C_p$ to become negative.  The following
well-known relation exists \cite{Reif}:
\begin{eqnarray}
C_p-C_V=VT\frac{\alpha^2}{\kappa}  \nonumber
\end{eqnarray}
where $\alpha$ is the volume coefficient expansion and $\kappa$ is the
isothermal compressibility given by
\begin{eqnarray}
\alpha &=& \frac{1}{V}(\frac{\partial V}{\partial T})_p \nonumber \\
\kappa &=& -\frac{1}{V}(\frac{\partial V}{\partial p})_p \nonumber
\end{eqnarray}
For negative $\kappa$, $C_p$ is less than $C_V$ and can become negative.

Using the equality $(\frac{\partial V}{\partial T})_p=-(\frac{\partial V}
{\partial p})_T(\frac{\partial p}{\partial T})_V$ we can also write
\begin{eqnarray}
C_p-C_V=T(\frac{\partial p}{\partial T})_V(\frac{\partial V}
{\partial T})_p
\end{eqnarray}
This shows that $C_p$ can drop below $C_V$ if the isobaric volume coefficient
of expansion becomes negative which is the case in some regions of fig. 11.

\vskip 0.1in
\begin{center}
\includegraphics[width=3.5in,height=3.5in,angle=-90]{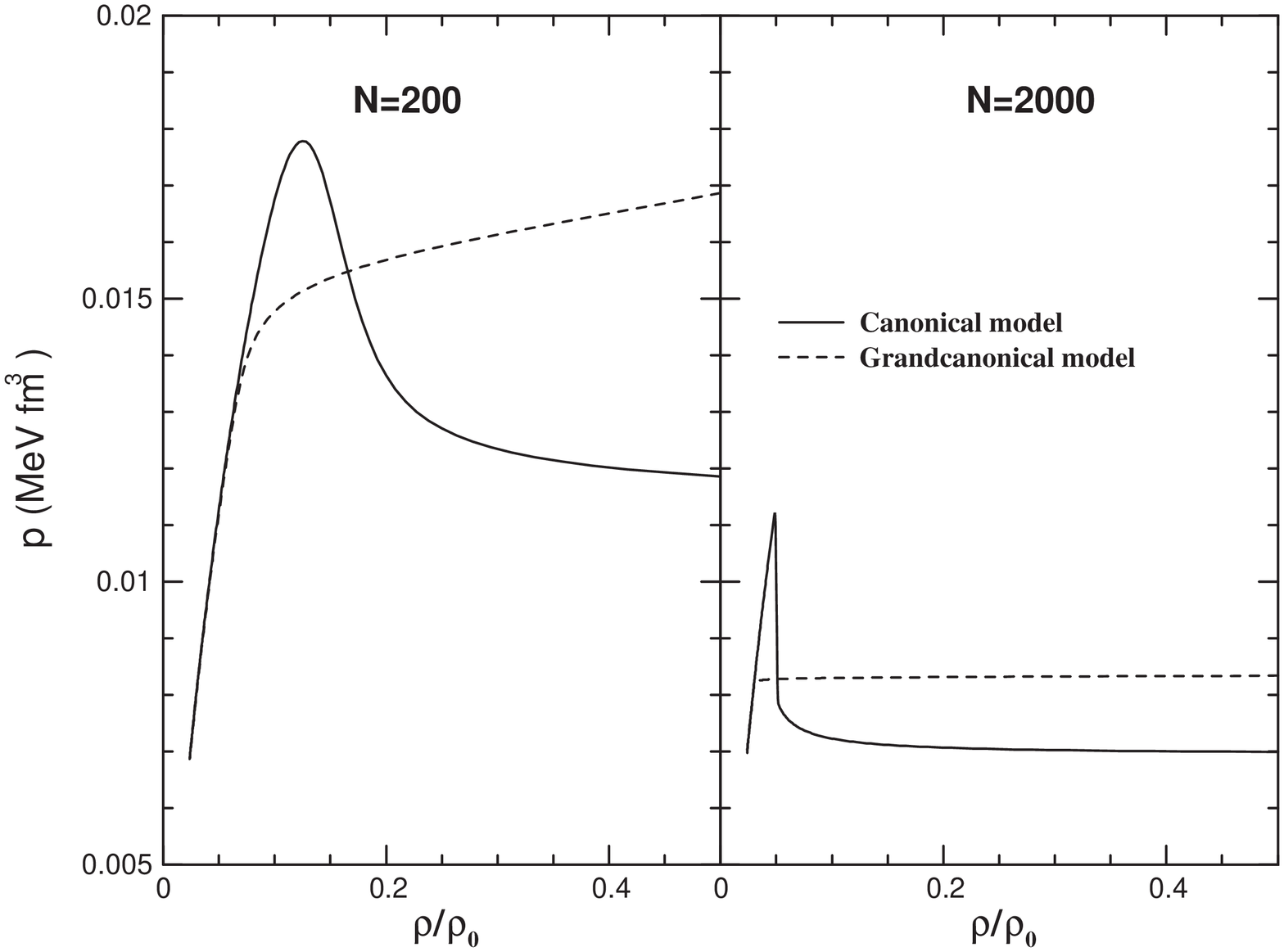}
\end{center}

\vskip 0.1in
Fig. 14.  EOS at $T$=6 MeV in the two models.  For the left panel, the
largest cluster has $N$=200 and for the right panel $N$=2000.  For the
canonical calculation, the left panel has $A$=200 and the right panel
has $A$=2000.  For the grand canonical calculation $A=\infty$.

\vskip 0.1in
We leave now general considerations of phase transitions, specific heat, 
caloric curves
etc. and explore the predictive powers of the canonical
thermodynamic model in producing detailed data
in heavy ion reactions.  Specifically we will investigate
how effective the canonical thermodynamic model is
in predicting isotopic yields in some specific reactions.  For this
we need to go beyond the production of hot fragments that the canonical
thermodynamics will give.  To obtain yields of specific final products,
we need to investigate how fragments at non-zero temperatures will
decay.  The next sections address this issue.

\section{Corrections for secondary decay}

The statistical multifragmentation model described above
calculates the properties of the collision averaged system that
can be approximated by an equilibrium ensemble. Ideally, one would
like to measure the properties of excited primary fragments after
emission in order to extract information about the collisions and
compared directly with the equilibrium predictions of the model
described in this report. However, the time scale of a nuclear
reaction ($10^{-20}$ s) is much shorter than the time scale for
particle detection ($10^{-9}$ s) . Before reaching the detectors,
most particles decay to stable isotopes in their ground states.
Thus before any model simulations can be compared to experimental
data, it is indispensable to have a model that simulates
sequential decays. This turns out to be not a simple task.
                                                                                
In this section, we follow the techniques of refs.
\cite{betty1,betty2} to calculate the secondary decay. We identify
some issues that can be accurately addressed and others that are
less controlled and may contribute uncertainties that influence
the final results. Later, we calculate the secondary decay of
excited nuclei predicted by the statistical multi-fragmentation
model and compare the final ground state yields to recent
measurements.

\subsection{Levels and level densities}
                                                                                
To calculate the secondary decay corrections, one must specify
both the high lying states that are mainly populated at freeze-out
and the lower lying states whose populations increase as these
excited nuclei decay towards the ground state nuclei that are
experimentally measured. In the previous sections, the discussion
was centered about the states that are populated at freeze-out.
While in principle all nuclear states may be involved at
freeze-out, the vast majority of fragments are excited to the
particle unbound continuum.
                                                                                
The level densities in the unbound continuum influence the overall
yield of unbound nuclei at freeze-out as well as the sequence and
the number of particle decays. In principle, interactions between
fragments and their surroundings modify the states and their
excitation energies. The vanishing of the surface tension
$\sigma(T)$ in the free energy expression at the critical
temperature $T_c=18$ MeV reflects such considerations. Few
experimental constraints on continuum level densities exist,
however, even when the nuclei are isolated. Thus, uncertainties in
the continuum level densities introduce uncertainties into the
calculated results.

Following ref. \cite{betty1}, we represent the continuum level
densities corresponding to the internal free energies in Eqs. 7
and 13 by the expression:
\begin{equation}
\rho_{SMM}(E^{*},J)=\rho_{SMM}(E^{*})f(J,\sigma).
\label{eq_rhosmmj}
\end{equation}
\noindent where
$\rho_{SMM}(E^{*})=\rho_{FG}(E^{*})\,e^{-b_{SMM}(a_{SMM}E^{*})^{3/2}}$,
$b_{SMM}=0.07 A^{-1.82 \left(1+\frac{A}{4500}\right)}$, $a_{SMM} =
\frac{A}{\epsilon_{0}}+\frac{5}{2}\sigma_{0}\frac{A^{2/3}}{T_{c}^{2}
}$, $J$ is the spin, $E^{*}$ is the excitation energy and $A$ is
the mass of the fragment. For light and medium mass nuclei,
$a_{SMM} \approx A/8$. Here,
\begin{eqnarray}
\rho_{FG}(E^{*})&=&\frac{a_{SMM}^{1/4}}{\sqrt{4\pi}(E^{*})^{3/4}}
\exp\left( 2\sqrt{a_{SMM}E^{*}}\right),
\\
f(J,\sigma ) &=&\frac{(2J+1)\exp [-(J+1/2)^{2}/2\sigma
^{2}]}{2\sigma ^{2}},
\\
\sigma ^{2} &\approx&0.0888\sqrt{A\cdot E^{*}/8)}A^{2/3},
\end{eqnarray}
and  $E^{*}$ and $Z$ are the excitation energy and charge of the
fragment. For further details, we refer the reader to ref. \cite{betty1}.
                                                                                
In contrast to the continuum level densities, the discrete level
densities need no corrections for the influence of interactions
because these levels become important only much later in the decay
after the fragments have decoupled from their surroundings. For
this purpose, we use the spectroscopic information of isolated
nuclei with $Z < 12$ where the information is available. For $12
\leq Z \leq 15$, low-lying states are not well identified
experimentally and a continuum approximation to the discrete level
density \cite{betty3} was used. For all fragments with $Z \leq 15$
and excitation energies between the domains of discrete and
continuum level densities, the level densities were smoothly
interpolated \cite{betty1}. 

Where the experimental information for nuclei with $Z \leq 15$ is
incomplete, values for the spin, isospin, and parity were chosen
randomly in the decay calculations as follows: spins of 0-4
(1/2-9/2) were assumed with equal probability for even-A (odd-A)
nuclei, parities were assumed to be odd or even with equal
probability, and isospins were assumed to be the same as the
isospin of the ground state. This simple assumption turns out to
be sufficient since most of spectroscopic information is known for
these low-lying states.
                                                                                
For excitation energies where little or no structure information
exists, levels were assumed to be specified by the relevant level
density expression. Groups of levels were binned together in
discrete excitation energy intervals of 1 MeV for $%
E^{\ast}<15$ MeV, 2 MeV for $15<E^{\ast}<30$ MeV, and 3 MeV for
$E^{\ast}>30$ MeV to reduce the computer memory
requirements. The results of the calculations do not appear to be
sensitive to these binning widths. A cutoff energy of
$E_{cutoff}^{\ast}/A=5$ MeV was introduced corresponding to a mean
lifetime of the continuum states at the cutoff energy about 125
fm/c. Where unknown, parities of these states were chosen to be
positive and negative with equal probability and isospins were
taken to be equal to the isospin of the ground state of the same
nucleus. In this fashion, a table of states for nuclei with $Z
\leq 15$ was constructed.

\subsection{Sequential decay algorithm}
                                                                                
Before sequential decay starts, hot fragments with $Z \leq 15$
were populated over the sampled levels in the prepared table
according to the
temperature. For the $i$th level of a given nucleus (A,Z) with its energy $%
E^{*}_{i}$ and spin $J_{i}$, the initial population is,
\begin{equation}  \label{eq_inipop}
Y_{i} =
Y_{0}(A,Z)\frac{(2J_{i}+1)\exp(-E^{*}_{i}/T)\rho(E^{*}_{i},J_{i})}
{ \sum_{i}(2J_{i}+1)\exp(-E^{*}_{i}/T)\rho(E^{*}_{i},J_{i})}
\end{equation}
where $Y_{0}$ is the primary yield summed over all states of
nucleus (A,Z) and T is the temperature associated with the
intrinsic excitation of the fragmenting system at breakup.

Finally all the fragments will decay sequentially through various
excited states of lighter nuclei down to the ground states of the
daughter decay products. The decay of fragments with $ Z
> 15$ was calculated according to the fission model of ref. \cite{betty4}.
The subsequent decay of excited fission fragments with $Z \leq 15$
was calculated according to the Hauser-Feshbach algorithm
described here. In this algorithm, eight decay branches of n, 2n,
p, 2p, d, t, $^{3}$He and alpha were considered for the particle
unstable decays of nuclei with Z$\leq $ 15. The decays of particle
stable excited states via gamma rays were also taken into account
for the sequential decay process and for the calculation of the
final ground state yields. If known, tabulated branching ratios
were used to describe the decay of particle unstable states. Where
such information was not available, the branching ratios were
calculated from the Hauser-Feshbach formula \cite{betty5},
                                                                                
\begin{equation}  \label{eq_branch}
\frac{\Gamma_{c}}{\Gamma} = \frac{G_{c}}{\sum_{d} G_{d}}
\end{equation}
where
\begin{eqnarray}  \label{eq_tcoeff}
G_d &=& \langle I_d I_e I_{d3} I_{e3}|I_p I_{p3}\rangle^2  \nonumber \\
&& \times \sum^{|J_d+J_e|}_{J=|J_d-J_e|}
\sum^{|J_p+J|}_{l=|J_p-J|} \frac{ 1+\pi_p \pi_d\pi_e
(-1)^l}{2}T_l(E)
\end{eqnarray}
for a given decay channel $d$ (or a given state of the daughter fragment). $%
J_{p} $, $J_{d}$, and $J_{e}$ are the spins of the parent,
daughter and emitted nuclei; $J$ and $l$ are the spin and orbital
angular momentum of the decay channel; $T_{l}(E)$ is the
transmission coefficient for the $l$th partial wave. The factor
$[1+\pi_{p}\pi_{d}\pi_{e} (-1)^{l}]/2$ enforces parity
conservation and depends on the parities $\pi=\pm1$ of the parent,
daughter and emitted nuclei. The Clebsch-Gordon coefficient involving $I_{p}$%
, $I_{d}$, and $I_{e}$, the isospins of the parent, daughter and
emitted nuclei, likewise allows one to take isospin conservation
into account.

For decays from empirical discrete states and $l\leq20$, the
transmission coefficients were interpolated from a set of
calculated optical model transmission coefficients; otherwise a
parameterization described in Ref. \cite{betty3} was applied.

\section{Comparisons to data}
Even though the structure of the low-lying states of the fragments
plays little role in properties of the hot system, these structure
effects become critical when the fragments cool later by secondary decay.
In the sequential decay algorithm described in the last section, in
addition to more sophisticated level densities, empirical binding
energies of the known nuclei are incorporated.  Where the empirical masses
are lacking, an improved mass formula \cite{Souza1,betty1} is employed.
To be self-consistent, the same masses and level densities are
used both in the thermodynamic model which produces the excited primary
fragments and in the subsequent sequential decay.  This self-consistency
requirement appears to be necessary \cite{Souza1} for some observables.
The resulting code which combines the thermodynamic model with the
sequential decay algorithm is referred to as ISMM for improved
Statistical Multifragmentation Model in the following sections of the 
report.                                                                      
          
To illustrate the capabilities of the thermodynamic model, we
calculate final ground state elemental and isotopic yields for
systems with $A_{0}=168$ and $Z_{0}=75$ and A$_{0}$=186 and
Z$_{0}$=75 at $T=4.7MeV$, corresponding to $E^{*}/A \approx 5$ MeV. 
In all the following
calculations, the freeze-out density is taken to be 1/6 of the
saturation density. These two systems were chosen
because they have the same proton fractions as
the combined systems formed in central $^{112}$Sn+$^{112}$Sn and
$^{124}$Sn+$^{124}$Sn collisions, respectively. However, the overall
size and excitation energy of these systems have been reduced
below that of the corresponding compound nuclei to reflect the
loss of particles and excitation energy to pre-equilibrium
emission prior to the multi-fragment breakup. These parameters
have not been adjusted to obtain a best fit of the data.

In the following, we illustrate the capability of this thermodynamic
model to describe experimental charge, mass and isotopic yield
distributions. We also compare experimental and calculated observables, such as
the isotopic temperature and the isoscaling parameters, which are
constructed from these yields.
                                                                                
\subsection{Charge and Mass Distributions}

Calculations of the mass distribution for excited
primary fragments are shown in Fig. 15 for a system
with $A_{0}=168$ and $Z_{0}=75$ at $T=4.7$ MeV. The distributions
of the primary fragments directly obtained from the thermodynamic model are
shown as dashed lines with open points while the solid 
line with solid points represent the
distributions of the final fragments after sequential decays.
Certain differences between primary and final spectra can be
expected. Heavier fragments formed in the multifragment stages
decay to smaller fragments, shifting the distribution to lower
masses. In addition, the decay produces a large increase in the
hydrogen and helium particles, because these are the main products
of the decay of the heavy fragments.
\vskip 0.1in
\begin{center}
\includegraphics[width=3.5in,height=3.5in,angle=90]{fig15.epsi}
\end{center}

\vskip 0.1in
Fig. 15.  Predicted mass distributions from the multifragmentation
of a source nucleus with the mass number 168 and charge number 75.
The open circles are primary yields and the closed circles are
yields after secondary decay.

\vskip 0.1in
The differential multiplicities $dM/d\Omega$ 
for various masses with $A \leq 20$ are plotted in an expanded
scale in Fig. 16 for both the $A_{0}=168$ and $A_{0}=186$ 
systems. For comparisons, experimental data obtained by averaging
over $70^{o} \leq \theta_{cm} \leq 110^{o}$ for central
$^{112}$Sn+$^{112}$Sn and $^{124}$Sn+$^{124}$Sn collisions at E/A=50
MeV \cite{betty6} are plotted as open and solid points in the left
and right panels, respectively. The calculations reproduce many
features of the mass distribution.

\vskip 0.1in
\begin{center}
\includegraphics[width=3.5in,height=4.5in,angle=90]{fig16.epsi}
\end{center}

\vskip 0.1in                                                                                
Fig. 16.  Predicted mass distributions ($A\leq 20$) from the 
multifragmentation of a source nucleus with $A_0$=168 and
$Z_0$=75 (left panel) and $A_0$=186 and $Z_0$=75.  The dashed
lines are the predicted primary yields and the solid lines are
predicted yields after secondary decay.  For comparison, data from
multifragmentation of central collisions of $^{112}$Sn+$^{112}$Sn
are shown as open symbols (left panel) and closed circles for
$^{124}$Sn+$^{124}$Sn reaction (right panel)[62].

\vskip 0.1in
The relative normalization of the calculation can be increased by
increasing the size of the source or by making its angular
distribution sideways peaked. The slope of the mass distribution
can be made more steep by increasing the source temperature. There
are indications that the experimental angular distributions are
not isotropic and that pre-equilibrium emission mechanisms may
contribute to the yields of the lighter fragments. Accordingly, we
do not fit the calculations to the experimental data in this
article, but defer such detailed analyses until more experimental
data that can constrain such effects become available.

The charge distributions exhibit similar behavior as the mass
distributions. For completeness, we include the charge
distributions for the $A_{0}=168$ and $Z_{0}=75$ and $A_{0}=186$
and $Z_{0}=75$ in Figs. 17 and 18.
The same conventions for the mass distribution figures 
(Fig 15 and 16) are used.

\vskip 0.1in
\begin{center}
\includegraphics[width=3.5in,height=3.5in,angle=90]{fig17.epsi}
\end{center}

\vskip 0.1in                                                                                
Fig. 17.  Predicted charge distributions from the multifragmentation
of a source nucleus with $A_0$=168 and $Z_0$=75.  The open circles
are primary fragment yields and the closed circles are after secondary
decay.

\vskip 0.1in                                                                                
\begin{center}
\includegraphics[width=3.5in,height=3.5in,angle=90]{fig18.epsi}
\end{center}

\vskip 0.1in                                                                                
Fig. 18.  Predicted charge distributions ($Z\leq 8$) from the
multifragmentation of source nuclei with $A_0$=168 and $Z_0$=75 (left panel)
and $A_0$=186 and $Z_0$=75.  The open and solid points are data from
Ref.[62].  See Fig. 16 for explanations of symbols used.

\vskip 0.1in                       
In the break up calculations, the odd-even effects are evident. These occur
because pairing and shell effects are not completely washed out in
our level density expressions at a temperature of $T = 4.7$ MeV.
As the secondary decay washes out such structures, these odd-even
effects in the primary distribution have little or no effect
on the final fragment distribution.
                                                           
\subsection{Isotopic distributions}
\begin{center}
\includegraphics[width=3.5in,height=3.5in,angle=90]{fig19.epsi}
\end{center}

\vskip 0.1in                                                                                
Fig. 19.  Isotope distributions for Carbon and Oxygen fragments.
The dashed and solid lines correspond to the predicted primary 
and final yields respectively.  The open and solid points are
from Ref. [62].

\vskip 0.1in                                                                                     
                                                                                
In Fig. 19, the isotopic distributions for carbon
and oxygen isotopes are plotted for the two sources. Using the
same convention as before, the dashed lines correspond to the
distributions of the primary fragments while the solid lines
represent the final distributions after sequential decay. 
As expected, the more neutron-rich system with $N_0/Z_0=1.48$
preferentially produces more neutron-rich isotopes than the
neutron deficient system with $N_0/Z_0=1.24$. 
In all
cases, the primary distributions are much wider and more
neutron-rich than the final distributions. The experimental
isotope distributions (data points) agree more with the final
results obtained after secondary decay than with the primary
distributions. Nonetheless, the widths of the experimental
distributions exceed those of the final distributions and are more
neutron-rich. This suggests that the predicted corrections for
secondary decay may be somewhat too large.

\vskip 0.1in
\begin{center}
\includegraphics[width=3.5in,height=3.5in,angle=90]{fig20.epsi}
\end{center}

\vskip 0.1in                                                                                
Fig. 20.  The mean neutron to proton ratios as a function of the
charge of the emitted fragment $Z$ for the two systems.  The left
and right panels correspond to the calculated results from the
primary and final fragments.

\vskip 0.1in                                                                                    
The mean neutron to
proton ratios $<N/Z>$ for each element provides another observable
with sensitivity to the isospin asymmetry dynamics of the
reaction. The dependence of the calculated primary values on the $<N/Z>$ of
the total system is much stronger than that of the final values.
This can be seen in Fig. 20 
where the primary (left panel) and final (right panel) $<N/Z>$ 
values are compared for the two systems. The
differences of the primary values for $<N/Z>$ of the two systems
are large, reflecting the large difference in the initial isospin
asymmetry of the two systems. The largest values for $<N/Z>$ occur
for $Z \approx 8, 20,$ etc., values corresponding to nuclei where
one can have either closed proton or neutron shells. Such
nuclei can remain comparatively well bound even for large value of
$N/Z$. Both of these enhancement and the difference between the
$<N/Z>$ values for the two systems are diminished in the final
distributions, which are both narrower and located closer to the
valley of beta stability.

Fig. 21 shows measured and calculated
primary and final values for $<N/Z>$ as functions of the element
number $Z$. The left and right hand panels provide the $<N/Z>$
values for the neutron-deficient and neutron-rich systems,
respectively. The calculated final distributions reproduce the
measured values well. It is rather curious that the 
experimental $<N/Z>$ values exhibit
the odd and even effects as a function Z. Such staggering is much
less obvious in the neutron rich system. For reference, the $<N/Z>$ 
for the abundances of naturally occurred isotopes are plotted as
stars in both panels of the figure.

\vskip 0.1in
\begin{center}
\includegraphics[width=3.5in,height=3.5in,angle=90]{fig21.epsi}
\end{center}

\vskip 0.1in                                                                                
Fig. 21.  The mean neutron to proton ratios as a function of the
charge of the emitted fragment $Z$ for the neutron deficient (left
panel) and neutron rich (right panel) systems.  For comparison, data
from the multifragmentation of central collisions of $^{112}$Sn+
$^{112}$Sn are shown as open symbols (left panel) and as closed circles
for $^{124}$Sn+$^{124}$Sn reaction (right panel)[62].  For reference,
the mean $N/Z$ ratios from naturally occurring isotopes are shown
as stars.

\vskip 0.1in                                                                                   
\subsection{Isoscaling}

The dependence of the isotopic distributions on the $N_0/Z_0$ of the
colliding system can be more sensitively explored by the use of
isotopic ratios \cite{betty6,betty7,betty8,betty9}. In particular, the
ratio, $R_{21}(N, Z) = Y_{2}(N, Z)/Y_{1}(N, Z)$, of yields
from two different reactions, labelled here as 1 and 2, has been
shown to exhibit an exponential relationship as a function of the
isotope neutron number $N$, and proton number, $Z$
\cite{betty6,betty7,betty8,betty9,betty10,betty11,betty12,betty13,betty14,betty15,betty16,betty17,betty18}.
                                                                                
\begin{equation}\label{}
  R_{21}(N, Z) = C \cdot exp(\alpha N + \beta Z)
\end{equation}
                                                                                
where $C$ is a normalization factor and $\alpha$ and $\beta$ are
the isoscaling parameters.

Calculations with a variety of different statistical models show
that the isoscaling relationship is strictly obeyed by the primary
fragments in these models \cite{betty8,betty10,betty15}.
Surprisingly the isoscaling relationship is also obeyed by
fragments produced in dynamical models such as the asymmetrized
molecular dynamical model \cite{betty14}. In all cases, the
isoscaling parameters are related to the isospin asymmetry of the
collisions and to the form of symmetry energy or, equivalently,
asymmetry term of the EOS chosen in the model
\cite{betty8,betty10,betty14,betty15,betty19}.
                                                                                
Neglecting for simplicity the Coulomb interactions between
fragments and environment, the exponential dependence of the
isoscaling relationship can be easily understood 
from the expression for the yields for a fragment with neutron and
proton numbers $N$ and $Z$ within the grand canonical limit of the
present equilibrium model \cite{betty20}:
                                                                                
\begin{equation}\label{}
Y_{i}(N,Z) = V_{i} \frac{A^{3/2}q _{N,Z}(T)}{\lambda
_{T_{i}}^{3}}\exp \left[ (Z\,\mu _{p,i}+N\,\mu _{n,i}
+B_{N,Z})/T\right].
\end{equation}

Here, $q _{N,Z}(T_{i})$ represents the internal partition function
of the fragment, $V_{i}$ the free volume of the system, $\lambda
_{T}=\sqrt{2\pi \hbar ^{2}/mT_{i}}$, $m$ the nucleon mass and $\mu
_{p_,i} $ ($\mu _{n,i}$)the chemical potential associated with free
protons (neutrons) for the $i^{th}$ reaction which produces a
system at temperature $T_{i}$. If the temperature in the two
reactions are expected to be the same (as in the Sn reactions
described here), the chemical potentials $\mu _{p_,i}$ and $\mu
_{n,i}$ contain the only reaction dependent factors in this
exponential. In this limit,$\alpha = [\mu _{n,2}-\mu _{n,1}]/T$
and $\beta = [\mu _{p,2}-\mu _{p,1}]/T$.
                                                                                
The symbols in Fig. 22 represent the isotopic
ratios calculated by the canonical thermodynamic model described in
this review. In Figs. 22 and 23, the following convention is
adopted. We choose closed symbols and solid lines for even Z and
open symbols and dashed lines for odd Z starting with Z=1 for
leftmost line. The lines are best fits of the calculated 
$R_{21}$ ratios to Eq. (33); the lines are essentially linear and
parallel on this semi-log plot consistent with a single constant
isoscaling parameter $\alpha_{primary} = 0.50$. The spacing between
these lines corresponds to the increase in $R_{21}$ for unit
increases in $Z$; the observed equal spacing is consistent with a
single constant isoscaling parameter $\beta_{primary} = -0.64$.

For comparison to the data, we only examine the isotope ratios where
there are data with sufficient statistics. The symbols in the
bottom panel of Fig. 23 represent the predicted
isotopic ratios after sequential decays. The lines are nearly parallel
to the lines in Fig. 22 on average and the
isoscaling parameters $\alpha_{final} = 0.46$ and $\beta_{final} =
-0.52$ are comparable to the primary values. In detail especially when the
isotopes away from the valley of stable nuclei are considered, the trends
are not as clearly consistent with the isoscaling law as are
the trends of the primary distribution. The larger change in the $\beta$ values
may arise from the approximation of the Coulomb interaction used in the
model. In the top panel, the data are shown as symbols. The experimental
isoscaling parameters are $\alpha_{data} = 0.36$ and $\beta_{data} =
-0.42$. The slopes from the calculations
are flatter suggesting that the temperature of 4.7 MeV used
as the input parameter in the model may be too low. However, if the 
temperature is increased so that the isoscaling predictions
agree with the data, the other observable such as the mass and charge
distributions as well as the isotope distributions may no longer agree. 
As stressed earlier,
the current work is not to use the optimized set of model parameters but
rather to compare the trends of data with the the model calculations.
More constraints and study are needed to optimize the agreement with data.

\vskip 0.1in
\begin{center}
\includegraphics[width=3.5in,height=3.5in,angle=90]{fig22.epsi}
\end{center}

Fig. 22.  Predicted yield ratios, $R_{21}(N,Z)=Y_2(N,Z)/Y_1(N,Z)$
from primary fragments for the two systems studied in this work.
The lines are best fit to the symbols according to Eq. (33).
Different lines correspond to $Z$=1 to 8 starting with the leftmost
line with three points being $Z$=1.

\vskip 0.1in                                                                                    
\begin{center}
\includegraphics[width=3.5in,height=3.5in,angle=90]{fig23.epsi}
\end{center}

\vskip 0.1in                                                                                
Fig. 23. Top Panel: Experimental isoscaling behavior
 exhibited by the central $^{112}$Sn+$^{112}$Sn and $^{124}$Sn+$^{124}$Sn 
collisions. The
 data are the nuclide yield ratios, $R_{21}(N,Z)$ from the two reactions
 plotted as a function of N. The isotopes of different elements lie 
along different lines. The solid and dashed lines represent the best
 fit to Eq. (33). Bottom Panel: Predicted yield ratios, $R_{21}(N, Z)$ 
obtained from the final yields for the two systems studied in this work.
 The symbols and lines have the same convention as the data used in the top 
panel and Fig. 22.
\vskip 0.1in 
\subsection{Isotopic temperatures}
                                                                                
Starting from the grand canonical expression for the yields (Eq.
34), it is also possible to construct a double ratio that
minimizes the sensitivity to the isospin asymmetry while
maximizing the sensitivity to the temperature. By doing so, one
can construct an isotopic thermometer, whereby the temperature is
extracted from a set of four isotopes produced in multifragment
breakups as follows \cite{betty21} ,
\begin{equation}
T_{iso}=\frac{\Delta B}{\ln (aR)}  \label{eq_albergo}
\end{equation}
where
\begin{eqnarray}
R
&=&\frac{Y(A_{1},Z_{1})/Y(A_{1}+1,Z_{1})}{Y(A_{2},Z_{2})/Y(A_{2}+1.Z_{2})},
\label{eq_tratio}\\
\Delta B &=&B(A_{1},Z_{1})-B(A_{1}+1,Z_{1})  \nonumber \\
&&-B(A_{2},Z_{2})+B(A_{2}+1,Z_{2}),
\end{eqnarray}
and
\begin{equation}
a=\frac{\left( 2J_{Z_2,A_2}+1\right) \left( 2J_{Z_1,A_1+1}+1\right)
}{\left( 2J_{Z_1,A_1}+1\right) \left( 2J_{Z_2,A_2+1}+1\right) }\left[
\frac{A_{2}\left( A_{1}+1\right) }{A_{1}\left( A_{2}+1\right)
}\right] ^{3/2}.
\end{equation}
                                                                                
In this ratio derived from Eq. (36) for the ground state
yields, $Y(A,Z)$ is the yield of a given fragment with mass A and charge Z; 
$B(A,Z)$ is the binding energy of this fragment; and $J_{Z,A}$ is
the ground state spin of the nucleus. In the context of the grand
canonical ensemble, Eq. (35) has been regarded as an effective or
\textquotedblleft apparent\textquotedblright\ temperature that may
differ somewhat from the true freezeout temperature $T$ due to the
influence of secondary decay and other cooling mechanisms.
                                                                                
The influence of secondary decay on the isotopic temperatures can
be clearly observed because it leads to variations in the values
for the temperature that depend on the isotopes used to construct
the ratio. The variations are universal, observed in many
different reaction systems and thus can be used to assess the
effectiveness of sequential decay models. One origin of these
variations is the feeding from higher lying particle bound states.
Such effects can be modeled by changing the value for the
statistical factor "a" and making it temperature dependent. This
and additional feeding from the decay of heavier particle unbound
nuclei can be modeled by the secondary decay formalism described
in the previous section.
                                                                                
To illustrate the influence of secondary decay on isotope
temperature measurements, measured and calculated final
temperatures have been extracted from double ratios of Z=2-8
fragments and plotted in Fig. 24. To reduce the
influence of secondary decay, we include only isotope thermometers
with large values for $\Delta B$ in this figure. This requirement
restricts comparisons to three types of thermometers: a.) $
T_{iso}(^{3,4}He)$ with $Z_{2}$=2, $A_{2}$=3, b.)
$T_{iso}(^{11,12}C)$ with $Z_{2}$ =6, $A_{2}$=11, and c.)
$T_{iso}(^{15,16}O)$ with $Z_{2}$=8, $A_{2}$=15. We note that the
thermometer (a) involves the light particle pair $^{3,4}$He while
thermometers (b) and (c) concern only intermediate mass fragments
with $Z=3-8$. The solid lines show corresponding ISMM predictions
for these three types of thermometers as a function of $A_{1}$.
                                                                                
Similarities in the variations of the calculated and measured
temperatures allow insight into their origin. Each panel of Fig.
24 corresponds to fixed values of $Z_{2}$ and
$A_{2}$; the observed variations in $ T_{iso}$ are therefore
correlated with $Z_1$ and $A_{1}$. The highest values for
$T_{iso}$ involve $^{10}$Be ($Z_1$=4, $A_{1}$=10) and $^{18}$O
($Z_1$=8, $A_{1}$=18). The calculations attribute this increase
to enhancements in the yields of these nuclei due to $\gamma$-ray
feeding from their many low lying particle bound states
\cite{betty24}. Other thermometers in Fig. 24 provide
temperature values that are significantly lower than those
involving $^{10}$Be and $^{18}$O. Most thermometers are
 significantly lower than the
primary temperature of 4.7 MeV, depicted by the horizontal 
dashed line in the three panels.

\vskip 0.1in
\begin{center}
\includegraphics[width=3.5in,height=3.5in,angle=90]{fig24.epsi}
\end{center}

Fig. 24.  Isotopic temperatures extracted from three types of
thermometers.  Experimental data are shown as symbols.
The lines are calculations.  For reference, the input primary
temperature of 4.7 MeV is shown as the horizontal dashed
lines.  (See text for details on the dot-dashed line in the
left panel.)

\vskip 0.1in                                                                                   
Both calculated and measured values display a Z or A dependence in
$T_{iso}$. Calculated values for $T_{iso}(^{15,16}O)$ are about
0.5 MeV lower than those for $ T_{iso}(^{11,12}C)$,  which are
about $~0.2$ MeV lower than $ T_{iso}(^{3,4}He)$. There is also a trend
for isotopic temperature values to decrease as a function of
$A_{1}$. The calculated decrease of $ T_{iso}$ with $A_{1}$ and
$A_{2}$ reflect the increasing importance of multi-step secondary
decay contributions to the yields of these heavier nuclei. Such
multi-step decays make the system appear cooler because the final
ground state nuclei originate from the decay of an ensemble of
unstable nuclei that are less excited than the original ensemble.

We note that the experimental $T_{iso}(^{3,4}He)$ temperatures
(solid symbols in the left panel) are systematically higher than
the corresponding ISMM values (solid line). As these
thermometers derive their sensitivity to the temperature from the
large binding energy difference between $^{3}$He and $^{4}$He, the
difficulty in reproducing these quantities may arise if there are
significant pre-equilibrium production mechanisms for light
particles such as $^{3}$He \cite{betty24}. To illustrate this
effect, we assumed that 2/3 of the measured $^{3}$He yield is of a
non-thermal origin. This increases the $^{3}$He yield by a factor
of three; calculations including this pre-equilibrium enhancement
are shown as the dot dashed line in the left panel. The success
of this resolution of the discrepancies between
$T_{iso}(^{3,4}He)$ and $T_{iso}(^{11,12}C)$ suggests that it may
be necessary to make careful estimations of the contributions from
pre-equilibrium emission before isotope temperature measurements
involving $T_{iso}(^{3,4}He)$ will be fully accurate.

\section{Summary}
The canonical version of the thermodynamic model has helped clarify
many aspects of intermediate energy heavy ion collisions.  The
obvious advantage is that, as opposed to the grand canonical model,
it has an exact number of particles.  The predictions of the grand
canonical model (which really applies to very large systems) can
differ very significantly from those of the canonical model
specially in the intermediate energy regime.  The canonical model
helps us to understand the order of phase transition, the caloric
curve and the possibility of negative specific heat.  The model
gives quantitative fits to experimental data on isotopic yields
and the phenomenon of isoscaling, now so well established in
intermediate energy heavy ion collisions.  The virtue of the
model is also its simplicity.  Most of the calculations reported
in this work can be carried out quite easily.

\section{Acknowledgement}
This work is supported by Natural Sciences and Enginnering Research
Council of Canada, by Quebec Department of Education, National Science
Foundation under Grant No. PHY-01-10253 and US Department of Energy
Grant No. DEFG02-96ER40987.

\appendix
\section{Equilibrium, reactions and reaction rate time scales}

A basic assumption of statistical models is that equilibrium is reached in
the time scale of the reaction. For fragment or composite particle
distributions a complex set of reactions takes places
\cite{Mekjian2,Mekjian3}.  The processes
involved in the collision of heavy ions can be modeled in a manner that is
similar to nucleosynthesis in a dense, heated and evolving system such as
in the expansion of the early universe and in supernovae explosions.  The
starting point of such a description is then a dense and heated system of
neutrons and protons which combine through a set of reactions to make the
composite nuclei from the lightest nuclei such as deuterons, alpha
particles,.. ,  all the way up to much heavier and complex nuclei. By way
of illustration and also for contrast, the nucleosynthesis in the early
universe occurs through a set of two body reactions with the first element of
the chain being an electromagnetic radiative capture of a neutron plus
proton
to a deuteron with an emitted photon carrying away the excess energy. After
this first electromagnetic process, light elements are produced by a
sequential set of two body reactions such as $d+d\rightarrow
He^3+n,\ d+d\rightarrow t+p,\ t+d\rightarrow
He^4+n,$...Nuclei up to $Li$ are believed to be produced at
their equilibrium concentration in big bang nucleosynthesis models. The
abundance of heavy elements comes from processes involved in supernovae.
The study of these processes is the area of nuclear astrophysics and heavy
ion collisions offer the opportunity to study similar processes and
phenomena in the laboratory.

In heavy ion collisions, electromagnetic processes are too slow over the
time scale of the collision to produce the observed distribution of
composites or produced particles.  A typical time scale of the collision is
$10^{-22}$ sec or 30 $fm/c$ which is much shorter than any electromagnetic
process time scale. Densities in heavy ion 
collisions can be high enough for a three
body process to occur such as $n+p+N\rightarrow d+N,$
where the nucleon $N$ can
carry away the excess energy. At very high energies, meson production
processes occur, so that a $d$ is formed in radiative pion emission of  $n+p$.
Heavier composite particles evolve through reactions such
as those listed above. However, it should be noted that because of possible
very high initial densities, multi body processes can occur besides two
body processes even for composites heavier than the deuteron. These
only enhance the approach to equilibrium.  At RHIC energies, particle
production becomes very important, and reactions leading to new particles
have been studied \cite{Mekjian3,Koch}.

As an example of a reaction rate approach consider the formation of a
deuteron through the process $p+n+N\rightarrow d+N$.
The time evolution of the deuteron density $\rho_d$ can be obtained
from an equation involving the proton
density $\rho_p$, neutron density $\rho_n$, and nucleon density $\rho_N$:
\begin{eqnarray}
\frac{d\rho_d}{dt} =[\rho_p\rho_n[(\frac{\rho_d}{\rho_n\rho_p})_{eq}-\rho_d]
\rho_N \langle \sigma[N+d\rightarrow n+p+N] \times v \rangle
\end{eqnarray}

Here $(\frac{\rho_d}{\rho_n\rho_p})_{eq}$ is the equilibrium ratio of the
densities of $d$'s to $n$'s
and $p$'s and is a function of temperature. The $\langle \rangle$
term in A.1 involves the
product of the breakup cross section of deuterons induced by nucleons and
$v$, which is the relative velocity of the $N$ and $d$ pair. This product is
averaged over the velocity distribution of the pair.  In obtaining the
expression in A.1 we used detailed balance which relates the forward rate
for the formation process $p+n+N\rightarrow d+N$ to the backward rate of
the break up or absorption process $d+N\rightarrow p+n+N$. Equilibrium is
reached when the forward
rate is equal to the absorption rate. Initially, the deuteron density is
being built up by forward processes which involves the product of proton,
neutron and nucleon densities, but later in this time evolution deuterons
will start to be absorbed by backward processes which involve the newly
formed deuterons and the existing nucleons. Once equilibrium is reached
these underlying processes vanish in the description of the deuteron
density, which is now described by phase space factors with temperature and
volume playing a dominant role. Large volumes reduce composites since
nucleons are less likely to be near each other to combine and high
temperatures increase break up probabilities. Binding energy terms appear
as Boltzmann factors and enhance composite densities.

We can question whether rates are fast enough to produce equilibrium
distributions.  To answer this question we consider the following
simplified expression for a reaction rate: $\rho_N \times \sigma \times v$.
For $\rho_N$ we take
nuclear matter density or 0.15 nucleons/fm$^3$. Typically temperatures are 10's
of MeV for medium energy collisions and a temperature of 10MeV has a
kinetic energy of $15 \ MeV= (1/2)m(v/c)^2$. For $v/c=1/5$, a cross section
$\approx$ 1$fm^2$
will have a rate $10^{22}$/s. The reciprocal of this rate is the reaction rate
time scale which is $10^{-22}$s. Thus, a cross section of 1$fm^2$ will have a
reaction rate time scale that is equal to the characteristic time scale of
the collision. Under these circumstances equilibrium will be reached.

Next, consider the prototype two body reaction $A+B\rightarrow C+D$.
The rate of
growth of the density of $C$ can be related to the chemical
activity $A=\mu_A+\mu_B-\mu_C-\mu_D$, where $\mu_A$ is the chemical
potential of $A$, etc. Specifically, the time evolution of the density
of  $C$ is
\begin{eqnarray}
\frac{d\rho_C}{dt} =\rho_A\rho_B \langle \sigma[A+B\rightarrow C+D] \times v \rangle
(1-\exp[-A/T])
\end{eqnarray}
At equilibrium $\mu_A + \mu_B = \mu_C + \mu_D$. Thus, the factor $(1-
\exp[-A/T])\rightarrow 0$.  Near equilibrium $A < < T$ and $(1- \exp[-A/T])
\rightarrow A/T$. In
this limit the reaction rate eq.(A.2) is linear in the chemical activity $A$.
Such linear connections are known as Onsager relations where the chemical
activity acts as a generalized force, $X$,  and the left hand side of
eq.(A.2) is interpreted as a generalized velocity $J$. Then $J=LX$, where $L$
is the proportionality constant between $J$ and $X$.
Far from equilibrium, this
linear relation is no longer valid since $A$ is, in general, not small
compared to $T$.

As a final consideration in discussing reaction rates we note that if the
equilibrium concentration of the particle of interest is small, then the
reaction rate constant is somewhat more complicated than the simplified
expression used above. To illustrate this situation we mention the case of
pion production. For example, for the reaction $N+N\rightarrow N+N+\pi$, the
rate equation for the pion density is:
\begin{eqnarray}
d(\rho_{\pi})/dt = [\rho_N^2 - (\rho_N^2*\rho_{\pi}/(\rho_{\pi})_{eq}]
\times \langle \sigma v \rangle.
\end{eqnarray}

Here,$ (\rho_{\pi})_{eq}$ is the equilibrium pion density
which depends on temperature.
This rate equation can be solved to give
$\rho_{\pi}(t)=(\rho_{\pi})_{eq}*(1-\exp[-\lambda \times t])$.
The rate constant is
\begin{eqnarray}
\lambda = \langle \sigma v \rangle \times \rho_N^2/(\rho_{\pi})_{eq}
\end{eqnarray}

The result of eq.(A.4) differs from the simplified reaction rate used above
by an important factor $\rho_N/(\rho_{\pi})_{eq}$. This
factor can be very large when the
equilibrium density of pions is small compared to the nucleon density. It
was one of the reasons why the results of \cite{Mekjian3} led to the
conclusions that pions would be in chemical equilibrium, a result which
differed from a previous result in \cite{Sobel}.
While low equilibrium concentration can enhance
reaction rate constants and reduce equilibration time scales, some examples
of other enhancement factors are the presence of two or more channels to
the final state, the presence of secondary processes, high densities which
allow multiparticle production processes above the two body type just
considered. For example, the time scale for kaon production is considerably
reduced through pion induced reactions, where the pions are copiously
produced in the initial nucleon-nucleon collisions as first noted in
\cite{Mekjian3}.

\section{Antisymmetry and all that}
Our whole discussion started from eq.(2) in section 2
which then led to eq.(5), the
recursive formula.  Eq.(2) is not quantum mechanical.  The partition
function of $n_i$ particles takes this simple form only under situation
of low density and high temperature.  We argue here that the approximation
is quite good for intermediate energy heavy ion collisions.

We start with qualitative arguments.  The volumes used are about three times
or more of the normal volume.  At low temperature ($\approx 4$ MeV)
where one might imagine the approximation to fail, it survives because
many composites appear thus there is not enough of any particular species
to make (anti)symmetrisation an important issue.
At much higher temperature the number
of protons and neutrons increase but as is well-known the $n!$ correction
takes the approximate partition function towards the proper one at high
temperature.  In a hypothetical world, the problem could get very
difficult.  Such a scenario would arise if the physics was such that at
low temperature we only had neutrons and protons and no composites.
An even worse situation would be if we had only neutrons (or protons).
With these preliminaries let us proceed to estimate quantitatively the
errors involved in actual cases that one might encounter in intermediate
energy heavy ion collisions.

The recursive relation eq. (5) is not limited to the approximation of
eq.(2).  It is shown in \cite{rab5} that by regarding the
grand partition function (in our case this grand partition function
incorporates correct (anti)symmetry among particles) as the generating
function of the canonical partition function one derives a relation like
eq.(5)
\begin{eqnarray}
Q_N(\beta)=\frac{1}{N}\sum_{k=1}^{N}kx_kQ_{N-k}(\beta)
\end{eqnarray}
where $x_k$ is not a one-particle partition function but is to be
obtained from an
expansion of the grand partition function.  We illustrate this
with first the example of only protons filling up orbitals $i,j,k,...$
in a box.  Now
\begin{eqnarray}
lnQ_{gr}(\beta,\mu) & = &\sum_i ln(1+e^{\beta\mu-\beta\epsilon_i}) \\
\nonumber
& = & \sum_i\sum_j\frac{(-)^{j-1}}{j}e^{j(\beta\mu-\beta\epsilon_i)}
\end{eqnarray}
The coefficient of $e^{\beta\mu k}$ is $x_k$ which
then gives $x_k=\frac{(-)^{k-1}}{k}\sum_ie^{-k\beta\epsilon_i}$
When this expression for $x_k$ is used in eq. B.1 it generates
the correct partition function.  Orbitals are given occupancies greater
than one and then eliminated by subtraction.  This can lead to
severe round-off errors when applied to degenerate Fermi systems
but will not affect the application we envisage here.  The number of
of protons is given by
\begin{eqnarray}
Z=(\frac{x_1Q_{Z-1}}{Q_Z}+\frac{2x_2Q_{Z-2}}{Q_{Z}}+...\frac{Zx_ZQ_0}{Q_Z})/Z
\end{eqnarray}
The value of $Q_0$ is 1.

Anticipating generalisation we will call $x_k$ in the above case
$y_{1,0}^{[k]}$.  The subscript $1,0$ means it is a ``composite'' with
one proton and no neutron.  The superscript $k$ means it is obtained
from the $k-th$ term in the expansion; $y_{1,0}^{[k]}$ will contribute
to $x_{k,0}$.

If instead we had a boson, deuterons for example, we would have
\begin{eqnarray}
ln[Q_{gr.can}(\beta,\mu_p,\mu_n)] & = & \sum_i-ln(1-e^{\beta\mu_p+\beta\mu_n}
e^{-\beta\epsilon_i}) \\
& = &\sum_i\sum_j\frac{1}{j}e^{j(\beta\mu_p+\beta\mu_n-\beta\epsilon_j)}
\end{eqnarray}
Thus in the case of deuterons $y_{1,1}^{[k]}$ (which would contribute to
$x_{k,k}$) is given by $\sum_i\frac{1}{k}e^{-k\beta\epsilon_i}$.

We can treat an assembly of protons, neutrons, deuterons, tritons...etc.
The recursive relation if the dissociating system has $Z$ protons
and $N$ neutrons is
\begin{eqnarray}
Q_{Z,N}=\frac{1}{Z}\sum_{i=1,Z,j=0,N}ix_{i,j}Q_{Z-i,N-j}
\end{eqnarray}
The average number of a composite with $i_1$ protons and $i_2$ neutrons
is given by
\begin{eqnarray}
<n_{i_1,i_2}>=y_{i_1,i_2}^{[1]}Q_{Z-i_1,N-i_2}/Q_{Z,N}+2y_{i_1,i_2}^{[2]}
Q_{Z-2i_1,N-2i_2}/Q_{Z,N}+...
\end{eqnarray}
Unless one is in an extreme degenerate fermi system, one can evaluate
the $y$ factors by replacing sums with integration.  For example,
$y_{1,0}^{[n]}=
\frac{(-)^{n-1}}{n}\sum_i e^{-n\beta\epsilon_i}$ where the sum is
replaced by
$\int e^{-n\beta\epsilon} g(\epsilon)d\epsilon
=2\frac{V}{h^3}(\frac{2\pi m}{n\beta})^{3/2}$.  Here
$V$ is the available volume.  We have included the
proton spin degeneracy; $m$ is the proton mass.
For the deuteron, $y_{1,1}^{[k]}=
\frac{1}{k}\int e^{-k\beta\epsilon}g(\epsilon)d\epsilon$.  This is
$3\times 2^{3/2}\frac{V}{h^3}(\frac{2\pi m}{\beta})^{3/2}\frac{e^{k\beta E_b}}
{k^{5/2}}$ where $E_b$ is the binding energy of the deuteron.  It is clear
how to compute contributions from other composites.

We test the accuracy of the yields as calculated throughout the main
text by comparing with a calculation where the complete theory of
symmetrisation and antisymmetrisation is used.  Subject only to the
approximation that summation over discrete states has been replaced by
an integration over a density of states, the calculation is exact.
The results are taken from \cite{rab5}.  We take the dissociating system
to have $Z$=25 and $N$=25.  The lowest temperature considered is 3
MeV (one might argue that at lower temperature a model of sequential
decay is more appropriate).  The highest temperature shown is 30 MeV.
We take a freeze-out volume in which the composites can move freely
as three times the volume of a normal nucleus with 50 nucleons.
Aside from neutrons and protons we allow the possibility of composites.
Excited states of the composites were not allowed (they could have been
included but the purpose of the exercise was to compare two models:
calculations without the inclusion of excited states were sufficient
to reach conclusions).  Spins and binding energies for deuteron, triton,
 $^3$He and $^4$He are taken from experiments.  For higher mass composites
the binding energy is taken from empirical mass formulas.  For fermions,
spin 1/2 was assumed and for bosons spin 0 was assumed.  For each $Z$ we
take $N=Z-1, Z,$ and $Z+1$.  We present in the table average yields of
protons, neutrons, tritons, $^3$He, $^4$He and the sum of yields of
all nuclei with charges greater than 12.  The temperature range of 3 to
6 MeV are of interest to many experiments.  We also show results
at 30 MeV.  The approximation used in the main part of the text
is seen to be quite good.

\begin{table}
\begin{center}
\caption{ \it Comparision of claculations of average yields and E/A. By exact
we mean a calculation with proper symmetry.  Sum over discrete orbitals
in a box has been replaced by integration as is the usual practice.}
\begin{tabular}{cccccccccc}
\hline
\multicolumn{1}{c}{Calc} &
\multicolumn{1}{c}{$p$} &
\multicolumn{1}{c}{$n$} &
\multicolumn{1}{c}{$d$} &
\multicolumn{1}{c}{$t$} &
\multicolumn{1}{c}{$^{3}He$} &
\multicolumn{1}{c}{$^{4}He$} &
\multicolumn{1}{c}{$Z>12$} &
\multicolumn{1}{c}{Temp.} &
\multicolumn{1}{c}{$E/A$} \\
\multicolumn{1}{c}{} &
\multicolumn{1}{c}{} &
\multicolumn{1}{c}{} &
\multicolumn{1}{c}{} &
\multicolumn{1}{c}{} &
\multicolumn{1}{c}{} &
\multicolumn{1}{c}{} &
\multicolumn{1}{c}{} &
\multicolumn{1}{c}{(MeV)} &
\multicolumn{1}{c}{(MeV)} \\
\hline
approx&0.307&0.032&0.050&0.007&0.054&0.679&0.945&3&-7.863 \\
exact&0.306&0.031&0.051&0.007&0.053&0.696&0.945&3&-7.861 \\
approx&1.174&0.898&1.177&0.560&0.641&2.489&0.051&6&-4.117 \\
exact&1.117&0.856&1.195&0.553&0.638&2.573&0.050&6&-4.135 \\
approx&4.127&3.955&4.812&2.099&2.052&1.985&0.000&12&4.401 \\
exact&3.860&3.696&4.941&2.090&2.051&2.021&0.000&12&4.308 \\
approx&10.937&10.893&7.664&1.686&1.650&0.379&0.000&30&28.914 \\
exact&10.512&10.468&7.885&1.732&1.696&0.395&0.000&30&28.844 \\
\hline
\end{tabular}
\end{center}
\end{table}

\section{Applications to other areas}

While the main emphasis of this report is on the thermodynamic model
for nuclear multifragmentation , the applications of the approach
developed in section 2 to other areas will be mentioned in this
appendix. In particular, many problems in statistical mechanics can be
reformulated in terms of equations 1-5 in that section. Each problem
has a different choice for the factor $\omega$ that appears in these
equations and a different interpretation of it within the general
structure of those equations. We will now illustrate these remarks
with some examples.

Let us consider the following parallel between multifragmentation and
permutations, which appear when Fermi-Dirac and Bose-Einstein
statistics are included into problems with identical particles. Any
permutation can be broken up into cycle classes and this cycle class
decomposition is the basis for this parallel. A given permutation of A
particles has a specific cycle class decomposition which specifies the
number of cycles of length $k$.  This number is similar to the number
of clusters of size $k$ in a fragmentation. Moreover, the same type of
sum rule holds as with clusters. That is, for any given permutation,
the total $A$ is equal to the sum of the cycle length times the number
of cycles of that length in that specific permutation. The canonical
partition function for non-interacting particles such as Fermi-Dirac
or Bose-Einstein particles in a box or in a one body potential well
such as a harmonic oscillator well has a form given by eq.2 in section
2 \cite{rab1,rab2,rab3}. For identical particles in a box of volume
$V$ and a system at temperature $T$, the $\omega$ weight factor for a
cycle of length $k$ is that of eq.6 with the $q_k = 1$ in that
equation for Bose-Einstein particles and $q_k= (-1)^{(k+1)}$ for
Fermi-Dirac particles. Once the canonical partition function is
obtained from the recurrence relation of eq.5, the thermodynamic free
energy $F$ can be calculated and all other thermodynamic quantities
also follow from $F$. For example, the pressure will have a form
involving an expansion in density and quantum volume which gives the
quantum corrections to the ideal gas law coming from the
symmetrization or anti-symmetrization of the particles. Bose-Einstein
particles in a laser trap which is taken as a harmonic oscillator well
have also been studied using this approach \cite{rab3}. Fermions in a
well can also be studied as mentioned in \cite{rab3} and an extended
discussion can be found in \cite{rab4}. Interactions can also be
included along with quantum statistics as shown in \cite{rab5}. Some
further observations regarding permutations are as follows. The result
of eq.4 gives the mean number of cycles of length $i$ in terms of the
ratio of the two partition functions $A-i$ and $A$, and the $\omega$
factor for that length. Near the Bose-Einstein condensation transition
long cycle lengths start to appear and this manifestation of the
transition is analogous to the appearance of large clusters around the
liquid gas phase transition.  The results of eq.4. give the
probability of a particular permutation, specified by its $n$ vector,
being present.  In RHIC collisions many pions are produced and the
application of the methods in sec.2 can also be given. For example
Bose-Einstein effects associated with thermal pions have been studied
in \cite{rab6,rab7}. For thermal pions at temperature $T$ in a volume
$V$ the cycle length $\omega$ factor of eq.6 is given by,
\begin{eqnarray}
\left(\frac{VT^3}{2\pi^2}\right) {\left(\frac{m}{T}\right)}^2 \left(\frac{1}{k^2}\right) K_2\left[k\frac{m}{T}\right]. \nonumber
\end{eqnarray}
Here, $m$ is the mass of the pion and $K_2$ is a
MacDonald function. The $\omega$ weight factor also appears in
expressions concerning the mean number of pions, its fluctuations, and
in higher moments of the pion probability distribution. Examples of
these connections are:

\begin{eqnarray}
\langle N \rangle &=& \sum k \omega_k \nonumber \\
\langle N^2 \rangle - \langle N \rangle ^2 &=& \sum k^2 \omega_k \\
\langle {(N - \langle N \rangle )}^3 \rangle &=& \sum k^3 \omega_k   . \nonumber
\end{eqnarray}

The sums that appear in eq.C.1 are over all $k$'s. Note that Poisson
statistics has only unit cycles, or $k=1$ only in the sums. Then
$\langle N^2 \rangle - {\langle N \rangle }^2 = \langle N \rangle$. 
The presence of cycles of length 2 and higher
cycles produces departures from Poisson statistics. An important
observation related to Poisson statistics comes from the fact that
coherent states have associated Poisson distributions. Moreover,
departures from Poisson statistics are associated with chaotic
emission processes. At high temperatures, Maxwell-Boltzmann statistics
apply which leads to Poisson statistics in statistical models. The
pion probability distribution for having $N$ pions is the ratio of the
canonical partition function for a system of size $N$ divide by the
grand canonical partition function. This probability was investigated
in \cite{rab6} for the case of $158 \ GeV \ Pb + Pb$ collisions where
it is shown to have a Gaussian shape with a width that is about $10\%$
larger than a Poisson distribution with the same mean number of
pions. Many other models of pion and, in general, particle
multiplicity distributions can be developed in a similar manner by
specifying another form for $\omega_k$. Once
$\omega_k$ is given, all quantities of interest follow.  The importance
of a phenomenological approach to multiparticle distributions, which
is based on known distributions from probability theory,  is shown in
\cite{rab8,rab9,rab10,rab11,rab12}. Moreover, a wide range of physical
processes can be accommodated
using such an approach. A specific and frequently used distribution is the
negative binomial distribution where $\omega_k=x \frac{t^k}{k}$. The symbol
$x$ is the
negative binomial parameter while $t$ is another parameter that is important
in fixing the mean number of pions and its variance:
$\langle N \rangle = x \frac{t}{1-t}$ and $\langle N^2 \rangle - \langle N \rangle^2 = \langle N \rangle (1 + (\langle N \rangle /x))$.
The generalized approach in \cite{rab6,rab7}
also includes several well known specific probability distributions as
special cases of a more general distribution. Here, we will just mention
a few examples of various phenomena that can be found in \cite{rab6,rab7} which
are as follows: (1) Emission from systems with a variable signal to noise
ratio, where the signal is related to a Poisson processes which may
originate from a coherent state and a noise level given by a negative
binomial distribution. (2) Field emission from Lorentzian line shapes
and its connection to a Feynman-Wilson gas \cite{rab13}. (3) Pion laser models
\cite{rab14,rab15} and the role of Bose-Einstein enhancement for a Poisson
emitting
source. (4) Multiparticle emission as a one dimensional random walk process
along a jet axis.  A reader interested in the  application of the methods
of sect.2 to multiparticle multiplicity distributions can find the details
and several other individual cases in \cite{rab6,rab7}. In a series of papers
\cite{rab1,rab2},
Hegyi has considered many interesting aspects of multiparticle production
and has also introduced a generalized distribution for its description.

Photon count distributions can also be developed using the approach
of section 2.  In fact, early models of pionic distributions \cite{rab8} coming
from nucleon-nucleon and nucleus-nucleus collisions were based on photon
count distributions \cite{rab8}. The laser distribution of \cite{rab8} is an
example
of a distribution which first appeared in quantum optics and was then
subsequently taken over into the area of particle production. Thermal
emission of photons have an $\omega_k$ factor that can be obtained as
the zero mass limit of the pion result given above; namely $\omega_k =
2V T^3/(\pi^2 k^4)$. An additional factor of 2 appears for the spin of
the photon.


\begin{thebibliography}{00}

\bibitem{Hubele} J. Hubele et al., Z. Phys. A 340 (1991) 263.

\bibitem{Reif} F. Reif, Fundamentals of statistical and thermal physics,
(McGraw-Hill, New York,1965)chap. 8.

\bibitem{Mekjian1} A. Z. Mekjian, Phys. Rev. Lett. 38 (1977) 640.

\bibitem{Gosset} J. Gosset, J. I. Kapusta and G. D. Westfall, Phys. Rev.
C 18 (1978) 844.

\bibitem{Dasgupta1} S. Das Gupta and A. Z. Mekjian, Phys. Rep 72 (1981)
131.

\bibitem{Randrup} J. Randrup and S. E. Koonin, Nucl. Phys. A 471 (1987)
355c.

\bibitem{Gross1} D. H. Gross and H. Massmann Nucl. Phys. A 471 (1987)
339c.

\bibitem{Bondorf} J. P. Bondorf, A. S. Botvina, A. S. Iljinov, I. N. Mishustin
and K. Sneppen, Phys. Rep. 257 (1995) 133.

\bibitem{Dasgupta2} S. Das Gupta and A. Z. Mekjian, Phys. Rev. C  57
(1998) 1361.

\bibitem{Jaqamann} H. R. Jaqamann, A. Z. Mekjian and L. Zamick, Phys. Rev. C
27 (1983) 2782.

\bibitem{Curtin} M. W. Curtin, H. Toki and D. K. Scott, Phys. Lett B 123
(1983) 289.

\bibitem{Bertsch1} G. F. Bertsch and P. J. Siemens, Physics Lett. B 126
(1983) 9.

\bibitem{Dasgupta3} S. Das Gupta, A. Z. Mekjian and M. B. Tsang, Advances in
Nuclear Physics 26 (2001) 89 (Kluwer Academic/Plenum Publishers, New York).

\bibitem{Bhat1} P. Bhattacharyya, S. Das Gupta and A. Z. Mekjian, Phys.
Rev. C 60 (1999) 054616.

\bibitem{Chase} K. C. Chase and A. Z. Mekjian, Phys. Rev. C 52 (1995)
R2339.

\bibitem{Souza1} S. R. Souza et al., Phys. Rev. C 67 (2003) 051602.

\bibitem{Moretto} L. G. Moretto et al., Phys. Rep. 287 (1997) 249.

\bibitem{rab5}
B. K. Jennings and S.Das Gupta, Phys. Rev. C 62(2000)014901

\bibitem{Bauer1} W. Bauer, Phys. Rev. C 38 (1988) 1297.

\bibitem{Campi1} X. Campi, Phys. Lett B 208 (1988) 351.

\bibitem{Pan1} J. Pan and S. Das Gupta, Phys. Rev. C 51 (1995) 1384.

\bibitem{Dasgupta4} S. Das Gupta, J. Pan, J. Kvasnikova, and C. Gale,
Nucl. Phys. A 621 (1997) 897.

\bibitem{Bugaev} K. A. Bugaev, M. I. Gorenstein, I. N. Mishustin and
W. Greiner, Phys. Rev. C 62 (2000) 044320.

\bibitem{Bondorf2} J. P. Bondorf, R. Donangelo, I. M. Mishustin and
H. Schulz, Nucl. Phys. A 444 (1985) 460.

\bibitem{Stauffer} D. Stauffer and A. Aharony, Introduction to Percolation
Theory(Taylor and Francis, Washington DC, 1992) ch.2.

\bibitem{Widom} B. Widom, J. Chem. Phys. 43 (1965) 3898.

\bibitem{Fisher} M. E. Fisher, Physics 3 (1965) 255.

\bibitem{Elliott1} J. B. Elliott et al.,Phys. Lett. B 381 (1996) 35.

\bibitem{Elliott2} J. B. Elliott et al., Phys. Lett. B 418 (1998) 34.

\bibitem{Scharenberg} R. P. Scharenberg et al., Phys. Rev. C 64 (2001) 054602.

\bibitem{Gulminelli1} F. Gulminelli and Ph. Chomaz, Phys. Rev. Lett.
82 (1999) 1402.

\bibitem{Das1} C. B. Das, S. Das Gupta and A. Majumder, Phys. Rev C 65
(2002) 034608.

\bibitem{Gilkes} M. L. Gilkes et al., Phys. Rev. Lett. 73 (1994) 1590.

\bibitem{Beaulieu} T. Lefort et al., Phys. Rev. C 64 (2001) 064603.

\bibitem{Ruangma} A. Ruangma et al.,Phys. Rev. C 66, (2002) 044603

\bibitem{Das4} C. B. Das et al., Phys. Rev. C 66 (2002) 044602.

\bibitem{Goodman} A. L. Goodman, J. I. Kapusta and A. Z. Mekjian,
Phys. Rev. C 30 (1984) 851.

\bibitem{Elliott3} J. B. Elliott et al., Phys. Rev. Lett. 85 (2000) 1194.

\bibitem{Finn} J. E. Finn et al., Phys. Rev. Lett. 49 (1982) 1321.

\bibitem{Hufner} J. Hufner and D. Mukhopadhyay,  Phys. Lett.
B173  (1986) 373.

\bibitem{Oddershede} L. Oddershede, P. Dimon and J. Bohr, Phys. Rev. Lett
71 (1993) 3107.

\bibitem{Pan2} J. Pan, S. Das Gupta and M. Grant, Phys. Rev. Lett.
80 (1998) 1182.

\bibitem{Chomaz1} Ph. Chomaz and F. Gulminelli, Phys. Lett. B 447 (1999) 221.

\bibitem{Dasgupta5} S. Das Gupta and S. K. Samaddar in ``Isospin Physics
in Heavy-Ion Collisions at Intermediate Energies'' edited by Bao-An Li
and W. Udo Schroder (Nova Science Publishers, Inc, Huntington, New York,
2001) ch.4.

\bibitem {Muller1} H. Muller and B. D. Serot, Phys. Rev. C 52 (1995) 2072.

\bibitem{Muller2} H. Muller and B. D. Serot in ``Isospin Physics
in Heavy-Ion Collisions at Intermediate Energies'' edited by Bao-An Li
and W. Udo Schroder (Nova Science Publishers, Inc, Huntington, New York,
2001) ch.3.

\bibitem{Bhat2} P. Bhattacharyya, S. Das Gupta and A. Z. Mekjian, Phys.
Rev. C 60 (1999) 064625.

\bibitem{Das2} C. B. Das, S.Das Gupta and A. Z. Mekjian, Phys. Rev. C 67
(2003) 064607.

\bibitem{Das3} C. B. Das, S. Das Gupta and A. Z. Mekjian, Phys. Rev. C 68
(2003) 014607.

\bibitem{Das5} C. B. Das, S. Das Gupta and A. Z. Mekjian, Phys. Rev. C 68
(2003) 031601(R).

\bibitem{Lee1} S. J. Lee and A. Z. Mekjian, Phys. Rev. C 63 (2001) 044605

\bibitem{Lee2} S. J. Lee and A. Z. Mekjian, Phys. Rev. C 68 (2003) 014608

\bibitem{Lee3} S. J. Lee and A. Z. Mekjian, Phys. Lett. B 580(2004)137


\bibitem{Gross2} D. H. Gross, Phys. Rep. 279 (1997) 119.

\bibitem{Chomaz2} P. Chomaz, V. Duflot and F. Gulminelli, Phys. Rev.
Lett. 85 (2000) 3587.

\bibitem {Moretto2} L. G. Moretto, J. B. Elliott, L. Phair, and
G. Wozniak, Phys. Rev. C 66 (2002) 041601(R).

\bibitem{betty1} W. P. Tan et al., Phys. Rev. C 68, (2003) 034609.
                                                                                
\bibitem{betty2} W.P. Tan, PhD Thesis, Michigan State University (2002).
                                                                                
\bibitem{betty3} Z. Chen and C. K. Gelbke, Phys. Rev. C 38, (1998) 2630-2639.

\bibitem{betty4} R.J. Charity et al., Nucl. Phys. A483, (1998) 371;
R.J. Charity, M. Korolija, D.G. Sarantites, and L.G. Sobotka, Phys. Rev. C 56, 
(1997) 873; R.J. Charity, ibid. 58, (1998) 1073.
                                                                                
\bibitem{betty5}
W. Hauser and H. Feshbach, Phys. Rev. 87, (1952) 366.
\bibitem{betty6}
T. X. Liu et al., Phys. Rev. C 69, (2004) 014603.
\bibitem{betty7}
 M. B. Tsang, W. A. Friedman, C. K. Gelbke, W. G. Lynch, G. Verde, and H. S. Xu
Phys. Rev. Lett. 86, 5023 (2001).
\bibitem{betty8}
M. B. Tsang et al., Phys. Rev. C 64, (2001) 054615.
\bibitem{betty9}
H. Johnston et al., Phys. Rev. C 56, (1997) 1972.
                                                                                
\bibitem{betty10}
D. V. Shetty et al., Phys. Rev. C 70, (2004) 011601.
\bibitem{betty11}
Y. G. Ma et al., Phys. Rev.  C 69, (2004) 064610.
                                                                                
\bibitem{betty12}
S. R. Souza, R. Donangelo, W. G. Lynch, W. P. Tan, and M. B.
Tsang, Phys. Rev. C 69, (2004) 031607.
                                                                                
\bibitem{betty13}
M. Veselsky, G. A. Souliotis, and S. J. Yennello, Phys. Rev. C 69,
(2004) 031602.
                                                                                
\bibitem{betty14}
A. Ono, P. Danielewicz, W. A. Friedman, W. G. Lynch, and M. B.
Tsang, Phys. Rev. C 68, (2003) 051601.
\bibitem{betty15}
A. S. Botvina, O. V. Lozhkin, and W. Trautmann, Phys. Rev. C 65,
(2002) 044610.
                                                                                
\bibitem{betty16}
M. B. Tsang et al., Phys. Rev. Lett. 92, (2004) 062701.
                                                                                
\bibitem{betty17}
H. S. Xu et al., Phys. Rev. Lett. 85, (2000) 716. 
                                                                                
\bibitem{betty18}
M. Veselsky, G. A. Souliotis, and M. Jandel, Phys. Rev. C 69,
(2004) 044607.
                                                                                
\bibitem{betty19}
W. A. Friedman, Phys. Rev. C 69, (2004) 031601.
                                                                                
\bibitem{betty20}
C. B. Das, S. Das Gupta, X. D. Liu, and M. B. Tsang Phys. Rev. C
64, (2001) 044608. 
                                                                                
\bibitem{betty21}
S. Albergo et al., Nuovo Cimento A89, (1985) 1.
                                                                                
\bibitem{betty22}
M. B. Tsang, W. G. Lynch, H. Xi, and W. A. Friedman Phys. Rev.
Lett. 78, (1997) 3836.
                                                                                
\bibitem{betty23}
H. Xi, W. G. Lynch, M. B. Tsang, W. A. Friedman, and D. Durand
Phys. Rev. C 59, (1999) 1567.
                                                                                
\bibitem{betty24}
H. Xi et al. Phys. Rev. C 57, (1998) R462.

\bibitem{Mekjian2} A. Z. Mekjian, Phys. Rev. C 17 (1978) 1051.

\bibitem{Mekjian3} A. Z. Mekjian, Nucl. Phys. A 384 (1982) 492.

\bibitem{Koch} P. Koch, B. Muller and J. Rafelski, Phys. Rep. 142 (1986) 167.

\bibitem{Sobel} M. Sobel, P. J. Siemens, J. P. Bondorf and H. A. Bethe,
Nucl. Phys. A 251 (1975) 502.

\bibitem{rab1}
K .C.Chase and A.Z.Mekjian,  Phys. Rev. C 49 (1994) 2164.

\bibitem{rab2}
A.Z.Mekjian and S.J.Lee, Phy. Rev. A 44 (1991) 6294.

\bibitem{rab3}
K.C.Chase, A.Z.Mekjian and L.Zamick, European Phys. J. B 8 (1999) 281.

\bibitem{rab4}
S.Pratt, PRL 84 (2000) 4255.

\bibitem{rab6}
A.Z.Mekjian, B.Schlei and D.Stottman, Phys. Rev. C 58 (1998) 3627.

\bibitem{rab7}
S.J.Lee and A.Z.Mekjian, Nucl. Phys. A 730 (2004) 514.

\bibitem{rab8}
P.Carruthers and C.C.Shih, Int. J. Mod. Phys. A2 (1987) 1447.

\bibitem{rab9}
I.M.Dremin and J.W.Gary, Phys. Rep. 349 (2001) 301.

\bibitem{rab10}
E.A.Wolf, I.M.Dremin and W.Kittel, Phys. Rep. 270 (1996) 1.

\bibitem{rab11}
P.Bozek,M.Ploszajczak and R.Botet, Phys. Rep. 252 (1995) 101.

\bibitem{rab12}
S.Hegyi, Phys. Lett. B 309 (1993) 443, B 318 (1993) 642, B 327 (1994) 171.

\bibitem{rab13}
A.Z.Mekjian, Phys. Rev. C 65 (2002) 014907.

\bibitem{rab14}
S.Pratt, Phys. Lett. B 301 (1993) 159.

\bibitem{rab15}
T.Csorgo and J.Zimanyi PRL 80 (1998) 916.

\bibitem{rab16}
J.Klauder and E.Sudarshan, Fundamentals of Quantum Optics, Benjamin,
N.Y.  1968.

\end{thebibliography}
\end{document}